\makeatletter
\@ifundefined{@parse@version@dash}{%
\def\@parse@version#1{\@parse@version@0#1}
\def\@parse@version@#1/#2/#3#4#5\@nil{%
\@parse@version@dash#1-#2-#3#4\@nil}
\def\@parse@version@dash#1-#2-#3#4#5\@nil{%
  \if\relax#2\relax\else#1\fi#2#3#4 }
}{}
\makeatother
\documentclass[%
 aip,
 amsmath,amssymb,
 preprint,%
]{revtex4-1}

\usepackage{graphicx}
\usepackage{dcolumn}
\usepackage{bm}

\usepackage[utf8]{inputenc}
\usepackage[T1]{fontenc}
\usepackage{mathptmx}
\usepackage{etoolbox}
\usepackage[subrefformat=parens]{subcaption}

\makeatletter
\def\@email#1#2{%
 \endgroup
 \patchcmd{\titleblock@produce}
  {\frontmatter@RRAPformat}
  {\frontmatter@RRAPformat{\produce@RRAP{*#1\href{mailto:#2}{#2}}}\frontmatter@RRAPformat}
  {}{}
}%
\makeatother
\begin{document}

\preprint{AIP/123-QED}

\title[]{Numerical analysis of the efficiency of face masks for preventing droplet airborne infections}
\author{Keiji Onishi}
 \email{keiji.onishi@a.riken.jp}
 \homepage{http://www.r-ccs.riken.jp}
 \affiliation{Center for Computational Science, RIKEN, Kobe, 650-0047, Japan}

\author{Akiyoshi Iida}%
 \affiliation{Department of Mechanical Engineering, Toyohashi University of Technology, Toyohashi, 441-8580, Japan}%

\author{Masashi Yamakawa}
\affiliation{Faculty of Mechanical Engineering, Kyoto Institute of Technology, Kyoto, 606-8585, Japan}

\author{Makoto Tsubokura}
 \altaffiliation[Also at ]{Department of Computational Science, Kobe University, Kobe, 657-8501, Japan}
 \affiliation{Center for Computational Science, RIKEN, Kobe, 650-0047, Japan}

\date{\today}

\maketitle 
\section*{Abstract} 
 In this study, the flow field around face masks was visualized and evaluated using computational fluid dynamics. The protective efficiency of face masks suppressing droplet infection owing to differences in the shape, medium, and doubling usage is predicted. Under the ongoing COVID-19 pandemic condition, many studies have been conducted to highlight that airborne transmission is the main transmission route. However, the virus infection prevention effect of face masks has not been sufficiently discussed, and thus remains as a controversial issue. Therefore, we aimed to provide a beneficial index for the society. The topology-free immersed boundary method, which is advantageous for complex shapes, was used to model the flow in the constriction area, including the contact surface between the face and mask. The jet formed from the oral cavity is guided to the outside through the surface of the mask and leaks from the gap between the face and mask. A Darcy-type model of porous media was used to model the flow resistance of masks. A random variable stochastic model was used to measure particle transmittance. We evaluated the differences in the amount of leakage and deposition of the droplets during exhalation and inhalation, depending on the differences in the conditions between the surgical and cloth masks owing to coughing and breathing. The obtained result could be attributed to the epidemiological measures adopted against both the reduction of the risk of infecting others by suppressing the exhalation, and the inhalation amounts.



\section{Introduction}

 Under the ongoing COVID-19 pandemic conditions, it is becoming clear that the infection route by droplet dispersion plays a dominant role. Hydrodynamic considerations are significant in exploring the spread mechanism of COVID-19, and computational fluid dynamics (CFD) is expected to be utilized in all physical aspects of droplet transmission, from the formation of droplets to transportation by airflow \cite{Mittal2020}. However, the number of studies on the efficiency of face masks for droplet infection is insufficient. In addition, some numerical indicators for clarifying the effectiveness of wearing masks would be beneficial for the society, which is still a controversial issue.

 Many visualized study results on the personal protective equipment (PPE), particularly face masks, have been reported over the last year. Verma et al. \cite{Verma2020a} demonstrated the effect of face masks on cough events exhaled from a mannequin model that simulated the human body. They proved that the reach of dispersive droplet plumes can be reduced to less than 8.33\% using handmade masks and commercially available cone-style masks. In addition, Verma et al. have explored face shields \cite{Verma2020b}. Aydin et al. \cite{Aydin2020} and Fischer et al. \cite{Fischer2020} investigated the effect of hand-made face masks made of various materials on reducing the number of droplets. They showed that face masks have a droplet dispersion suppressing efficiency of approximately 70-90\%. Moreover, Stutt et al. \cite{Stutt2020} introduced an epidemiological mathematical model and a conventional susceptible infected removed compartmental model to predict the number of deaths from infection in the United States, which is consistent with their assumption that masks have an infection control effect of 50\% or higher. We presented a comparable model using the statistical data on a cohort of infected people in Japan, which revealed the same preventive effect of face masks that approximately 75\% of the population wore. However, the minimum particle size that can be captured by optical measurement is limited by the employed equipment and measurement conditions. The number of aerosol microdroplets contained in the human exhaled breath of several micrometers or less is controversial. For example, there are various theories regarding the number of droplets generated by human coughing. The number of particles of size $1 \mu m$ or less varies between 1,000 and 1,000,000 \cite{Duguid1946,Yang2007,Gupta2009,Zayas2012}. Although it has been reported that aerosol infection is the main transmission route for COVID-19, it is still unclear how much the infection risk changes with the number or volume of droplets inhaled. It is ideal to evaluate the number of droplets that cause infection; however, the basic behavior of the scattered droplets has not been sufficiently explored. Therefore, clarifying the behavior of droplets is the first step in understanding the phenomenon.

 In the numerical analysis of droplets discharged from the human exhale, some studies consider how far the droplets on the background airflow reach. Dbouk et al. \cite{Dbouk2020} have shown that generally accepted social distances of 2 m and 6 ft may not be sufficient for a single human coughing event. Vuorinen, et al. \cite{Vuorinen2020} calculated the detailed droplet dispersion conditions in supermarkets. It has been shown that the particles can reach farther distances depending on the background airflow conditions. Numerical analysis by ANSYS Inc. \cite{ANSYS2020} indicates that the reach of particles on the flow may change depending on the standing position of the marathon runners. The airflow and behavior of the droplets were not uniform and were significantly influenced by the environment. They shared the technique of expressing the flow field in the Euler grid system and coupling the motion of particles in the Lagrange system. Based on the multiphase analysis of fluid and solid particles, the details such as strong or weak coupling, and handling of humidity, temperature, and molar mass fraction are different. While the flow environment that must be reproduced is complex, the particle size is limited to small ones on the order of micrometers to millimeters; therefore, this approach seems to be the most efficient and reliable method at present.

 However, there are few examples of CFD studies on face masks. Lei et al. \cite{Lei2013} and Zhu et al. \cite{Zhu2016} investigated the splash-suppressing effect of N95 masks used by medical professionals. They concluded that in N95 masks, the resistance coefficient and expiratory flow velocity of the mask material have a direct effect on the filter-to-face seal elimination. It was assumed that the mask was in a close contact with the face. Nevertheless, their results cannot be referred to as a general conclusion for generally available face masks, such as disposable surgical masks, fabric masks, and handmade cloth masks. Because the N95 mask makes it easy to express the gap between the face and mask with a layer grid system, whereas typical surgical, fabric, and cloth masks have creases on, the flow path between the face and mask becomes complicated. There is a problem in reproducing the gap geometry, which makes it difficult to create a calculation grid. Inevitably, this gap has a significant influence on the droplet collection efficiency of the mask. Moreover, correct reproduction of this complicated flow path is the key to predicting the effect of a general face mask.

 In this study, a three-dimensional (3D) flow-droplet multiphase flow analysis was performed when wearing a face mask using a topology-free immersed-boundary method. The jet formed from the oral cavity is guided to the outside through the mask surface as a resistor and the gap between the face and mask. Threfore, it exhibited a complicated flow appearance. A porous Darcy-type model was used for the mask-resistance modeling. A probability droplet penetration model based on random numbers was used to model the particle transmittance. The immersed boundary method (IBM), which is advantageous for complex shapes, was used to calculate the constricted area, including the contact surface between the face and mask. Subsequently, the difference in the amount of droplet leakage for various masks and respiratory conditions during exhalation and inhalation were evaluated. From the enumerated results, we hope to contribute to epidemiological understanding against the reduction of the risk of infection from others to oneself by suppressing the number of droplets.

\section{Numerical methods}

 The governing equations are the spatially filtered incompressible Navier--Stokes and continuity equations, including the external forcing term of the IBM, denoted by $\bm{f}$. This formulation is used for large eddy simulation. The fundamental equations are non-dimensionalized using the Reynolds number $Re$, which are described as
\begin{align}
  \label{eq:mass}
  & \nabla \cdot \bar{\bm{u}} = 0, & \\
  \label{eq:momentum}
  & \frac{\partial \bar{\bm{u}}}{\partial t} + \bar{\bm{u}} \cdot \nabla \bar{\bm{u}} = - \nabla p + \frac{1}{Re} \nabla^2 \bar{\bm{u}} - \nabla \cdot \bm{\tau} + \bm{f}, & \\
\label{eq:stress}
& \tau_{ij} = \overline{u_{i} u_{j}} - \bar{u_{i}} \bar{u_{j}}, &
\end{align}
where $(\;\bar{ }\;)$ denotes the grid-filtering operator, $\bar{\bm{u}}$ is the filtered fluid velocity, $p$ is the pressure, and $t$ is the time. $\bm{\tau}$ is the subgrid-scale(SGS) stress tensor, which is generally modeled with $\bar{\bm{u}}$ to close the equation. The SGS models used in this study are based on the eddy viscosity concept that can be expressed as
\begin{align}
  \label{eq:SGSstress}
  & R_{ij} = \tau_{ij} - \frac{1}{3} \delta_{ij} \tau_{kk} = - \frac{2}{Re_{t}} \bar{\bm{S}}, \quad \bar{\bm{S}}=\frac{1}{2} \left( \nabla \bar{\bm{u}} + \left( \nabla \bar{\bm{u}}\right)^{T} \right), &
\end{align}
where $Re_{t}$ is the Reynolds number based on the eddy viscosity $\nu_{t}$, $\bar{\bm{S}}$ is the velocity strain tensor, and $\delta_{ij}$ is the Kronecker delta. The SGS model is defined as follows:

\begin{align}
  \label{eq:Smagorinsky}
  & \nu_{t} = C \bar{\Delta}^{2} |\bar{\bm{S}}|, \quad |\bar{\bm{S}}| = \sqrt{S_{ij} S_{ij}}, &
\end{align}
where $\bar{\Delta} = \left( \bar{\Delta_{1}} \bar{\Delta_{2}} \bar{\Delta_{3}} \right)^{1/3}$ is the filter width, provided by the grid width $\bar{\Delta_{i}}$ in the $i$-th direction.

\subsection{Turbulence model}

 The coherent structure model (CSM), derived by Kobayashi \cite{Kobayashi2005}, was used to model the eddy viscosity. This model is used for dynamically determining the local constant of the Smagorinsky SGS model based on the relationship between the second invariant of the velocity gradient tensor and the energy dissipation rate of vortices. It is advantageous to solve complex shapes, and it is designed to maintain simple treatment at a low computational cost.

 In CSM, the model coefficient $C$ is determined as follows. The second invariant of the velocity gradient tensor $Q$ is represented as:
\begin{align}
  \label{eq:Q}
  & Q = \frac{1}{2} \left( W_{ij} W_{ij} - S_{ij} S_{ij} \right), & \\
  & \bar{\bm{W}}=\frac{1}{2} \left( \nabla \bar{\bm{u}} - \left( \nabla \bar{\bm{u}}\right)^{T} \right), \quad |\bar{\bm{W}}| = \sqrt{W_{ij} W_{ij}}, &
\end{align}
where $\bar{\bm{W}}$ is the vorticity tensor. This definition is termed as the Q-criterion, which is often used to visualize the vortex structures. Kobayashi independently determined the model coefficients (in non-rotating flow) of the flow properties and Reynolds number as follows:
\begin{align}
  \label{eq:FCS}
  & F_{CS} = Q / E, & \\
  & E = \frac{1}{2} \left( W_{ij} W_{ij} + S_{ij} S_{ij} \right), & \\
  & C = C_{1} |F_{CS}|^{3/2} , \quad C_{1} = 1 / 20. &
\end{align}
Therefore, an eddy viscosity model is expressed, in which the coefficient is automatically zero in laminar flow, and is naturally damped in the wall direction, as discussed by Kobayashi.

\subsection{Topology-free immersed boundary method}

 The IBM forcing term is simply written as follows:
\begin{align}
  \label{eq:forcing_term}
  & \bm{f}^{n+1/2} =
  \begin{cases}
    \displaystyle -RHS^{n+1/2} + \frac {\bm{u}_{IB}^{n+1/2} - \bm{u}^n} {\Delta t}, & on \; immersed \; boundary, \\
    \displaystyle 0, & elsewhere,
  \end{cases}, & \\
  & \bm{u}=\bm{u}_{IB}^{n+1/2}, &
\end{align}
where $\bm{u}_{IB}^{n+1/2}$ is the known velocity associated with the boundary conditions on the immersed surface, and $RHS$ contains advection and diffusion terms in the cell centers near the immersed boundary.

 The advection-diffusion term was discretized using the second-order central difference scheme with 10\% blending of the QUICK scheme, the Crank Nicholson implicit method was used for the time progression, and the pressure-velocity correction was applied as the fractional step method. The red-black-colored successive over-relaxation method (SOR) was adopted as the solution method of the pressure Poisson equation. The pressure and velocity variables were located in the center of the cell.

 In IBM, the distribution function has a weighting function corresponding to the interaction with particles in the Lagrange system by the distance between the constituent points of the object and computational grid points in the Euler system. It is expressed using the delta function and applied as a correction term of the momentum equation for each calculation grid point. In topology-free IBM \cite{Onishi2021}, the existence of an object is implemented by dividing it into a fluid region $\Omega_{f}$, solid region $\Omega_{s}$, and boundary region $\Omega_{IB}$. The distribution function $\mathcal{D}$ is expressed in Eq. \ref{eq:distribution_function}. For simplicity, all interpolations are defined as linear functions of distance.

\begin{align}
  \label{eq:discrete_forcing_term}
  \mathcal{F}(\bm{x},t) &= \sum_{k} \bm{f}_{k}^{n+1/2} \delta (\mid \bm{x} - \bm{X}(s,t) \mid) & \nonumber \\
  & \approx \sum_{k} \bm{f}_{k}^{n+1/2} \mathcal{D} (\mid \bm{x} - \bm{X}(s,t) \mid), \quad \forall \: \bm{X}(s,t) \in \Gamma, &
\end{align}

\begin{align}
  \label{eq:distribution_function}
  & \mathcal{D} (\mid x - X \mid) =
  \begin{cases}
  \displaystyle \frac{\mid x - X \mid}{\Delta x + \mid x - X \mid}, & \quad for \; x \in \; \Omega_{IB}, \\
  \displaystyle \frac{\Delta x - \mid x - X \mid}{\Delta x + \mid x - X \mid}, & \quad for \; x \in \; \Omega_{s}, \\
  \displaystyle 0, & \quad for \; x \in \; \Omega_{f},
  \end{cases} &
\end{align}

where $\bm{X}(s,t)$ is a vector function that provides the location of points of the immersed boundary $\Gamma$ as a function of position $s$ and time $t$. In addition, $\bm{x}$ is the position of each cell center. $\delta$ is the Dirac delta function, which distributes the force imposed on the momentum equations of the surrounding cells. Moreover, this expression is a mixed usage of the Peskin-type IBM \cite{Peskin1972} and Mittal-type IBM \cite{Mittal2007}. The correction of momentum in the solid region is applied to dummy cells arranged in the virtual space for each cell. Solid cells with an arbitrary width can always be placed for fluid cells that pass through narrow gaps, and calculations can be made reasonably even for thin surfaces, intricate flow paths, and flow paths that have gaps/overwraps. The typical size of a dummy cell was $5 \times 5 \times 5$ cells. Furthermore, if the width of the distribution function is limited to one cell, the correction of the fluid region can be limited to the relevant fluid cell being calculated.

\begin{figure}[htb]
  \centering
  \begin{minipage}{0.3\hsize}
    \vspace{0.8cm}
    \includegraphics[keepaspectratio,width=\textwidth]{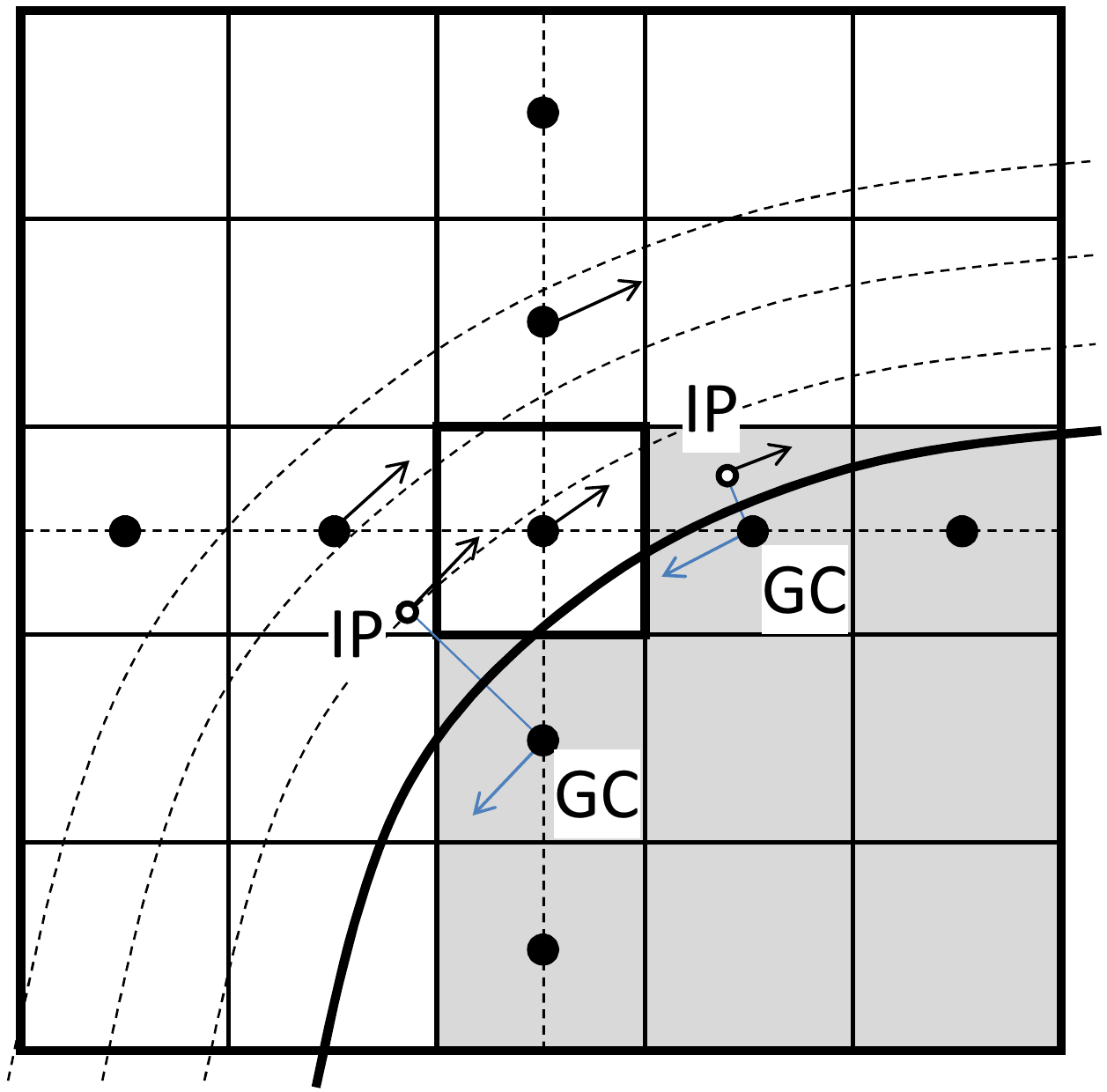}
    \subcaption{Ghost cell and imaginary point in Mittal's IBM.}
    \label{fig:wall_boundary_1}
  \end{minipage}
  \begin{minipage}{0.305\hsize}
    \includegraphics[keepaspectratio,width=\textwidth]{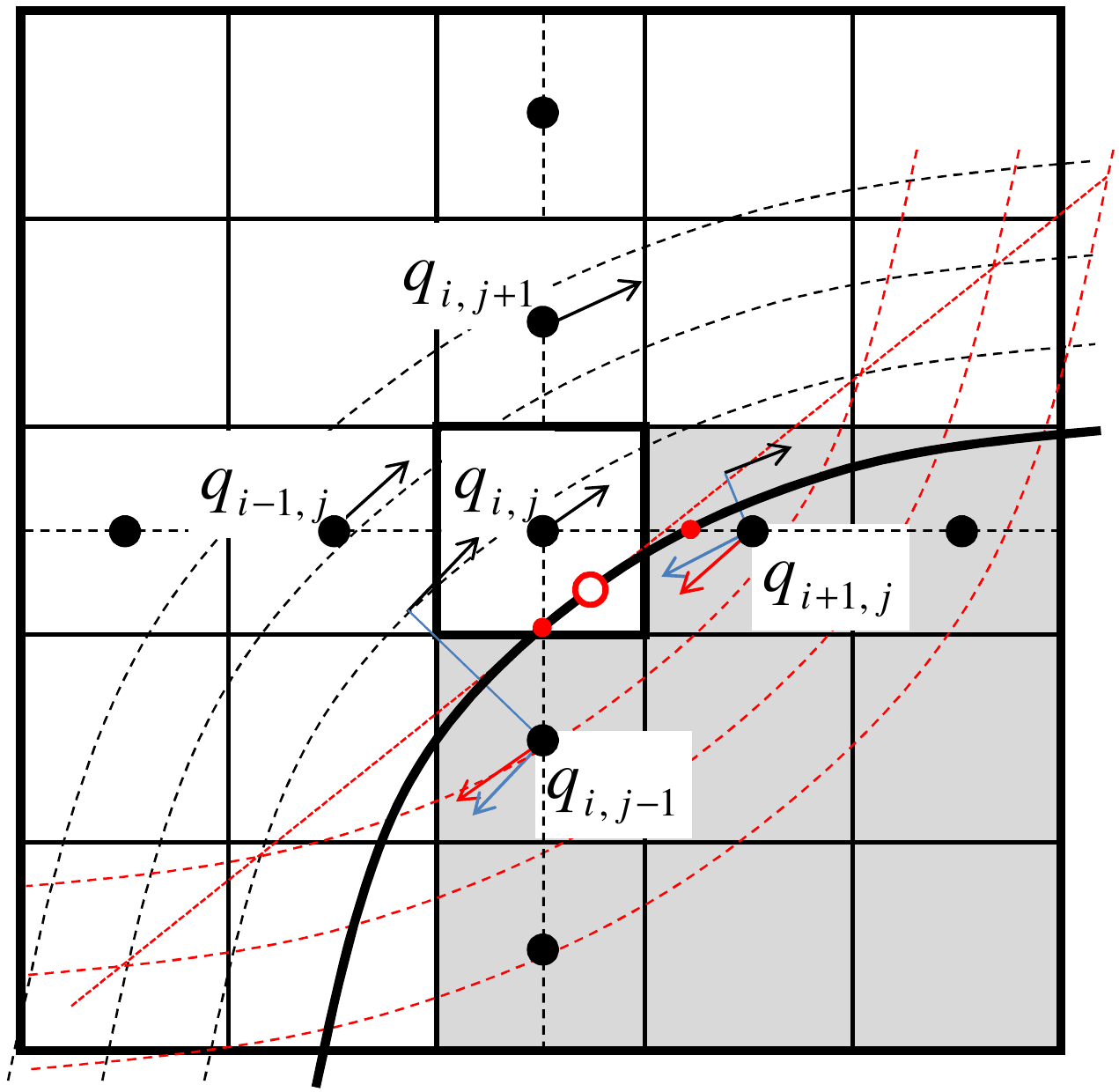}
    \subcaption{Axis-projected IBM.}
    \label{fig:wall_boundary_2}
  \end{minipage}
  \caption{Two-dimensional schematic view of the definition of the IBM wall boundary condition on the dummy cell. GC indicates a ghost-cell center, while IP indicates an imaginary point. The black dashed lines indicate flow field functions, while the red dashed lines indicate the sign-inverted flow-field functions. The red circle marks the intersection point of the perpendicular line from the cell center to the surface.}
  \label{fig:wall_boundary}
\end{figure}

 The modified momentum of the solid region was obtained by interpolation from the ambient fluid velocity field. The point at which the foot of the perpendicular line, drawn from the solid region cell to the wall surface, is extended by the same distance to the fluid region. The point where the interpolation occurs is called the IP or imaginary point. Subsequently, the obtained velocity is inverted and input to the solid cell (ghost cell, GC). The corrected momentum on the ghost cell is used only when the divergence of the equation of motion is calculated. Therefore, an additional transformation of interpolation is adopted. The interpolation is projected in the axial direction by satisfying the divergence form of the interpolation. Figure \ref{fig:wall_boundary} shows a conceptual diagram, where $q(i, j)$ represents the physical property of each grid point projected in the axial direction. This simplifies the calculation and avoids searching errors that occur in areas where it is difficult to judge fluids and solids (such as intricate shapes). By this transformation, the solution becomes robust and does not cause calculation errors regardless of the shape complexity. However, the calculation accuracy near the wall drops to the first-order accuracy as the grid resolution increases because it cannot be completely replaced in theory, as presented by Onishi et al. \cite{Onishi2021}.

 Using this formulation, the calculation becomes stable and satisfies the mass conservation for any shape topology. All GCs are well-posed, irrespective of the shape complexity, and there is no instability owing to the small cut cell or small distance. Moreover, this solution does not require an iterative method or uncertain constants. This method converged rapidly, even when the SOR was employed to solve the entire linear equation system. In addition, it is easy to parallelize the code.

\subsection{Lagrangian spray particle model}

 The equation of the motion of particles, known as the Maxey-Riley equation, assuming a sphere in the region where the Stokes approximation holds, was adopted as follows:

\begin{align}
  \label{eq:particle_motion}
  m_p (\frac{\mathrm{d} v_i}{\mathrm{d} t}) = \frac{1}{8} C_D  \pi d^2 \rho_{f} | u_i - v_i | ( u_i - v_i ) + m_p g_i, \\
  C_D = \frac{1}{24} ( 1 + 0.167 Re_{p}^{0.667}),
\end{align}

where $Re_{p}=|u_i-v_i|d/\nu$ is the particle Reynolds number, which was approximately 1 in this study. $m_p$ is the particle density (water), $u_i$ is the particle velocity, $v_i$ is the fluid velocity, and $\nu$ is the kinematic viscosity of the fluid. The motion of particles is calculated in each time step in a one-way coupling from the flow, without considering any feedback from particles to the fluid.
 In the interaction between the particle and wall, the ray-tracing algorithm is locally calculated to determine the tolerance between the particle trajectory and wall surface. Particles that interfere with the wall surface within a time increment are judged to move from two forms: adsorption (stick) and passage (penetration). Complete absorption is considered when it interferes with a solid wall. When it interferes with a face mask, adsorption or passage is selected from the probability of passage using a random generator. Because the measured gross value is used to model the mask transmission, detailed intermediate morphology of particle motion (rebound, break-up, splash, spread) associated with the wall surface interference are not considered in this study. Further, for the sake of simplicity, the effect of gravity is studied only after the droplets exit the mouth position.

\subsection{Face mask model}

 It is approximately impossible to calculate the face mask by 3D modeling of all the fibers contained in the material, even with the latest available supercomputers, owing to the requirement of extremely fine grid resolution, that is the order of $\mu m$. Therefore, we decided to replace it with a mathematical model that uses the pressure drop function and droplet transmittance.
 To model the permeation pressure drop of face masks, the Darcy-Forchheimer type resistance formula was used, which is commonly used in porous media flow physics. The resistance momentum is input as the external force term of the momentum equation.

\begin{align}
  \label{eq:porous_resistance}
  \nabla p= - \chi ( \alpha_{i} | \bm{u} | + \beta_{i} ) \cdot \bm{u}.
\end{align}

\begin{figure}[htb]
  \centering
  \begin{minipage}{0.45\hsize}
    \includegraphics[keepaspectratio,width=\textwidth]{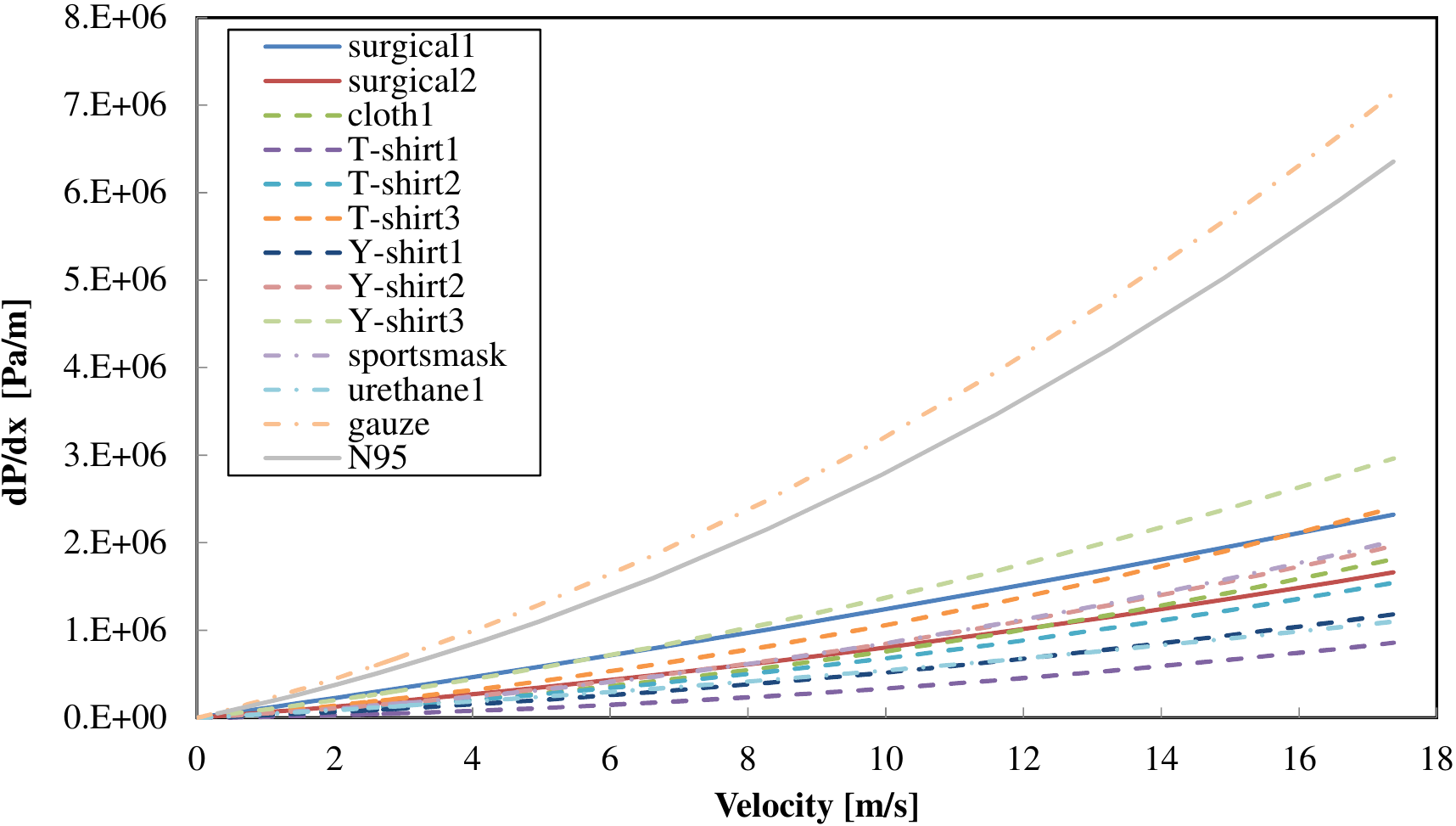}
    \subcaption{Pressure resistance of masks according to the flow velocity.}
    \label{fig:mask_resistance}
  \end{minipage}
  \begin{minipage}{0.51\hsize}
    \includegraphics[keepaspectratio,width=\textwidth]{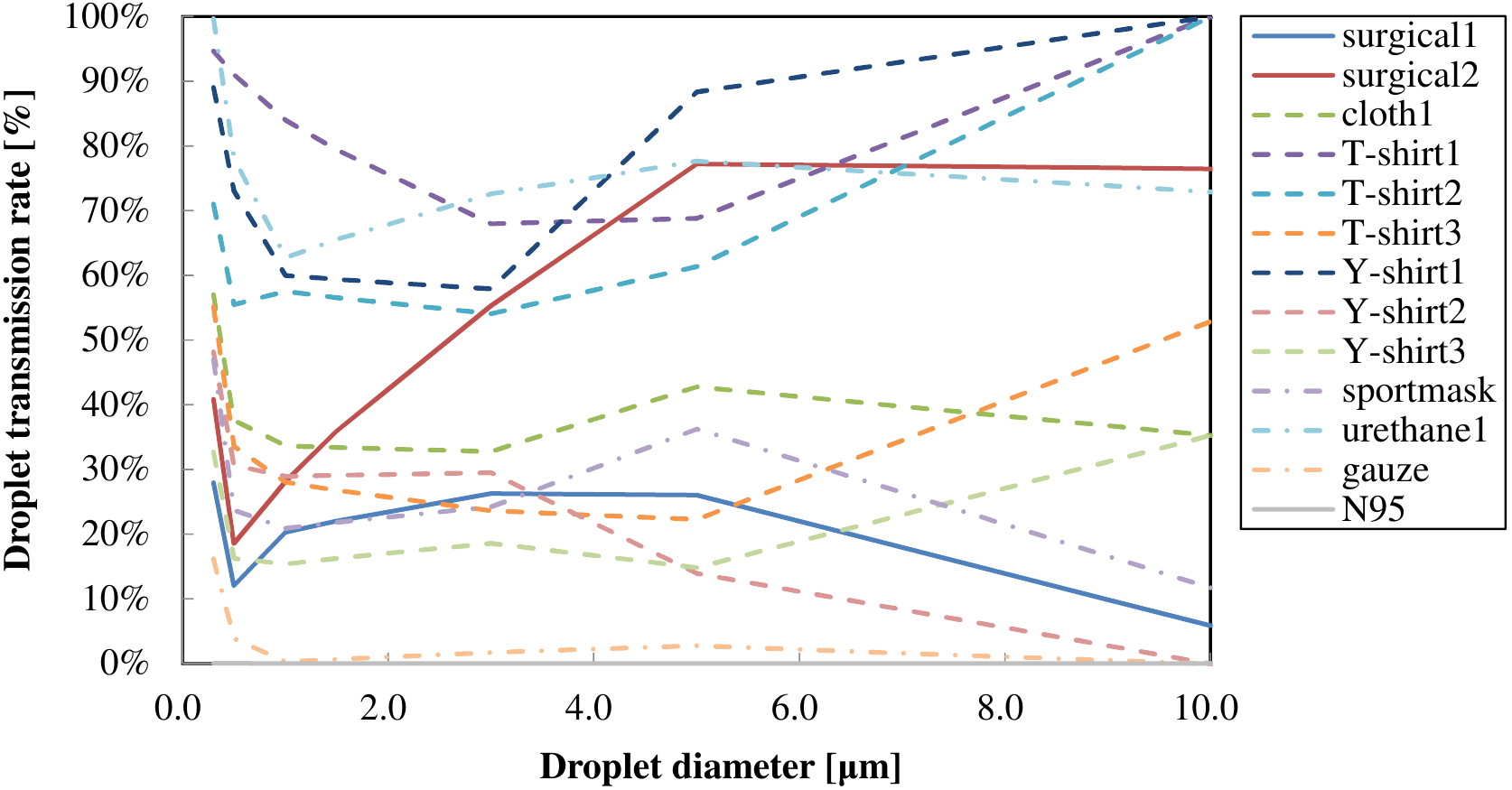}
    \subcaption{Droplet transmittance of masks according to the diameter.}
    \label{fig:mask_transmittance}
  \end{minipage}
  \caption{Mask resistance and transmittance parameters for each material. The mask thickness is $2.5 mm$ for all models.}
  \label{fig:mask_model}
\end{figure}

This is a model in which the pressure loss per unit length is proportional to the square of the velocity, and the coefficients $\alpha$ and $\beta$ are calculated using the experimentally measured values. Figure \ref{fig:mask_model}\subref{fig:mask_resistance} shows the resistivity of some materials according to the flow speed. In the calculation, the mask is implemented in a thickened shape (approximately $2.5 mm$) for all models. The external force term was calculated from the volume fraction of the cut cell. The direction of the external force was calculated using the local curvature of the face mask. A random variable stochastic model was used to calculate the transmittance of each particle. The transmittance for each particle size of the face mask filter was obtained using the results presented in the existing studies \cite{Sergey2009,Aydin2020} and our experimental results. Subsequently, the probability of penetrating the mask surface for each particle was calculated. Figure \ref{fig:mask_model}\subref{fig:mask_transmittance} shows the transmittance of some materials according to particle size. According to the experiments, the particle transmittance depends on the flow velocity. However, the difference in the results caused by the velocity dependence is not discussed in this study.

\subsection{Calculation conditions}

\subsubsection{Numerical setup for exhalation}\label{numerical_method_exhalation}

 Face masks are designed to be worn at all times as a protective device against exposure to dust, pollen, fine particle substances (e.g. PM2.5), and viruses. In addition, it serves as an anti-spreading filter that prevents cold viruses from spreading. The human body exhales particles while breathing, speaking, coughing, and sneezing. The number of particles released during a one-minute conversation is approximately 10,000, although the number varies according to the literature. The louder the conversation, the more particles tend to be included, as the affricate sound is included. Drinking alcohol naturally causes louder conversation, the amount of saliva secreted by eating and drinking increases, and the number of particles tends to increase further. The collection rate is desirable to be calculated by combining all of these conditions to evaluate the performance of a general face mask; however, this would increase the overall simulation cost. Therefore, we decided to focus on one typical condition, that is the coughing condition.

\begin{figure}[htb]
  \centering
  \begin{minipage}{0.4\hsize}
    \includegraphics[keepaspectratio,width=\textwidth]{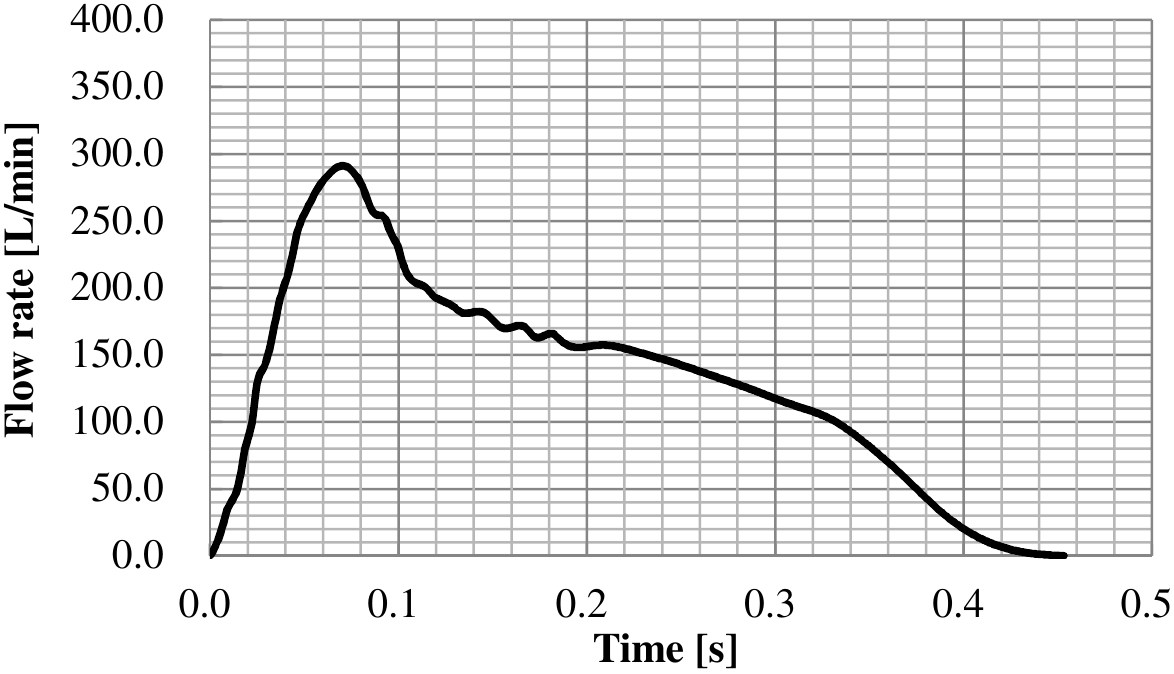}
    \subcaption{Flow rate of a single cough.}
    \label{fig:cough_flowrate}
  \end{minipage}
  \begin{minipage}{0.4\hsize}
    \vspace{0.1cm}
    \includegraphics[keepaspectratio,width=\textwidth]{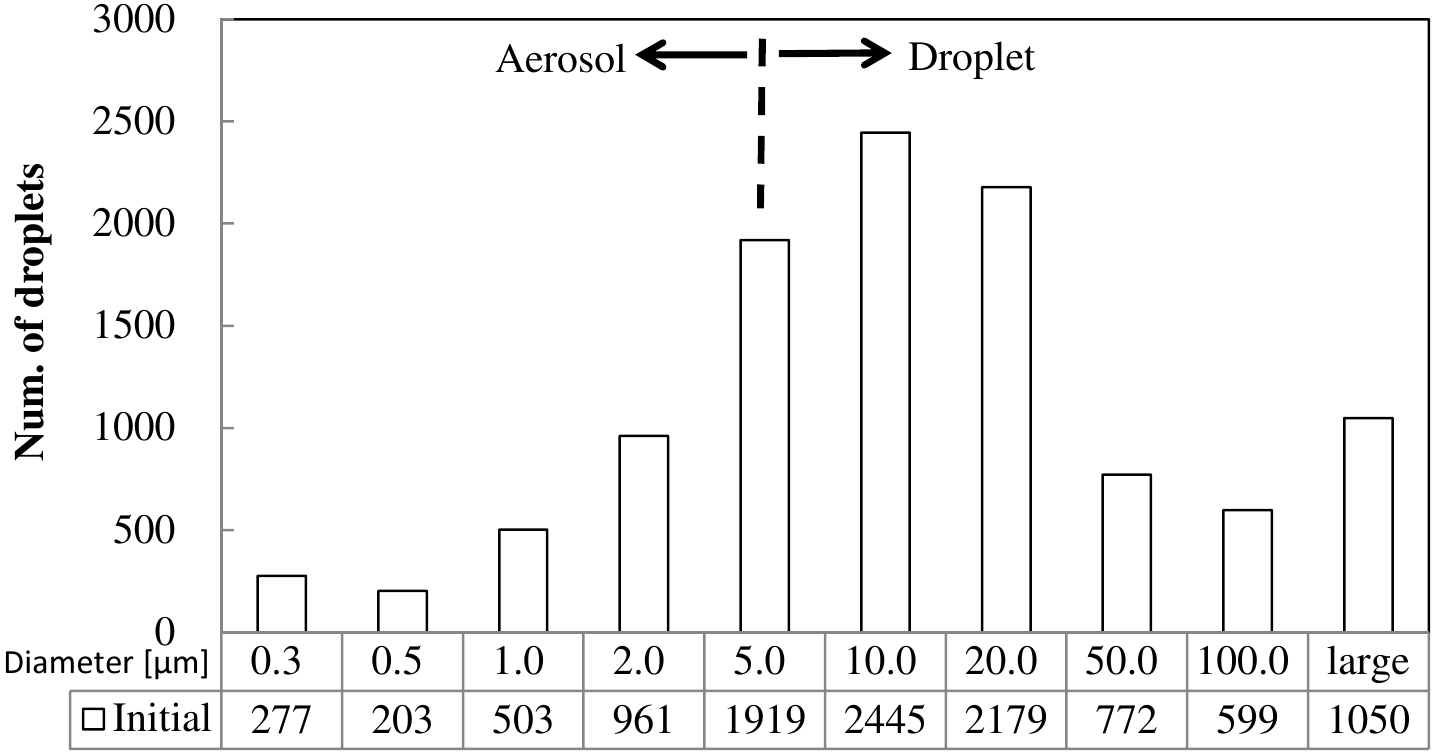}
    \subcaption{Number of droplets of a single cough.}
    \label{fig:cough_droplets}
  \end{minipage}
  \caption{Single cough conditions.}
  \label{fig:cough_model}
\end{figure}

 The flow rate of a single cough was collected from an existing study \cite{Gupta2009}, which is a time-series event with a peak value of approximately $292 L/min.$ for about $0.5 s$. The flow forms a cough cloud from the mouth opening and reaches a distance of approximately $2 m$ within $1 s$. The number of particles to create a distribution covering up to a large particle size was presented in some studies \cite{Duguid1946,Yang2007}. The total number of particles expressed by one cough was approximately 10,000. The particle size was randomized so that a smooth distribution could be obtained, with a peak of 10 $\mu m$, as shown in Fig.\ref{fig:cough_model}.

 The head geometry of the human was created using a 3D model \cite{CGTrader}, designed for computer graphics and distributed on the Internet. The size of the head to the top of the neck is approximately $ 237 \times 171 \times 247 mm $, which is close to that of a white adult male. The inflow channel from the throat to the mouth has a simple $3 cm \times 2 cm$ rectangle cross-section tubular elbow shape with respect to the experimental equipment, and the inflow boundary condition was set at the lower most end. The opening was smoothly joined to the 3D model of the head. As for the face mask shape, the finite element structural analysis was performed using a commercial software \cite{Wada2021}, and the solution that fitted the head model was solely obtained and used as a basis of the geometry calculations. The gap between the face and nose was approximately $2.9 cm^2$, and the gap between the face and cheeks was approximately $2.2 cm^2$ (one side). In addition, the effects of face mask deformation over time may be considered; however, it is omitted in this study for the sake of maintaining the problem simple.

\begin{figure}[htb]
  \centering
  \includegraphics[keepaspectratio,width=0.4\textwidth]{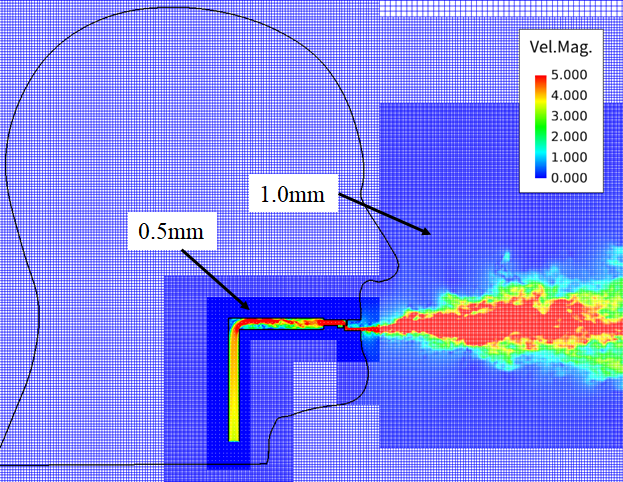}
  \caption{Grid resolution overview, colored by the flow velocity magnitude at a constant momentum flux input.}
  \label{fig:overview_flow}
\end{figure}

 The Reynolds number based on the peak velocity of one cough ($8.0 m/s $) and the opening height was $1.1 \times 10^{4}$. The flow distribution immediately downstream of the mouth opening was compared with the measured value using a hot-wire measurement. Thus, a qualitatively reasonable flow velocity distribution was obtained. The finest grid resolution was approximately $0.5 mm$ at the opening/inflow channel region, and approximately $1 mm$ near the face mask. Figure \ref{fig:overview_flow} shows an example of a grid image in the central section. Using the results obtained in our previous study \cite{Lu2021}, approximately the same solution could be obtained at a grid size of $2 \times$, where the resolution of the gap was not established at a grid size of $4 \times$, and a large number of calculations became difficult at a grid size of $1/2 \times$. The number of calculation grids included 4,089 cubes and 16.7 m cells. The time increment was set at $1.0 \times 10^{-5} s$ defined from the Courant condition of particles motion. The entire solution time was $1.0 s$ with $100,000$ time steps. The calculation time was approximately $2 h$ with 43 computing nodes (Fujitsu A64FX 2.0 GHz 48 cores, Fugaku) and approximately $11.5 h$ with 16 nodes (Intel Xeon 2.6 GHz 12 cores, Haswell).
 
\subsubsection{Numerical setup for inhalation}

 For an effective application of face masks to detain droplets during human breathing, the inhalation flow rate equivalent to deep breathing was appropriate \cite{Fenn1965}. The duration of an inhale, exhale, and another inhale was $6 s$. For the initial particle distribution, particles with a size in the range of 0-10 $\mu m$ were randomly arranged in an area of $r = 0.2 m$ from the center point of the opening for the fluid region outside the mouth/mask. The preventive effect was evaluated based on the number of inhaled particles.

 The shape of the inflow channel, including the larynx, pharynx, oral cavity, and nasal cavity, was constructed using the shape data of a real person (Japanese adult male) measured using the 3D scanning technology. It was smoothly merged with the 3D head model. The face mask shape uses the same model as in the exhale evaluation. The finest grid resolution was approximately $0.5 mm$ at the opening/inflow channel region, and approximately $1 mm$ near the face mask. The number of calculation grids was 8,338 cubes and 34.2 m cells. The time increment was set to $1.0 \times 10^{-5} s$. The entire solution time was $6.0 s$ with $600,000$ time steps. The calculation time was approximately $18.4 h$ with 82 computing nodes (Fugaku) and approximately $11.6 h$ with 32 nodes (Intel Xeon 2.7 GHz 56 cores, Cascade Lake).



\section{Results and discussion}

\subsection{Exhalation block efficiency evaluation}

\begin{figure}[htb]
  \centering
  \begin{minipage}{0.3\hsize}
    \includegraphics[keepaspectratio,width=\textwidth]{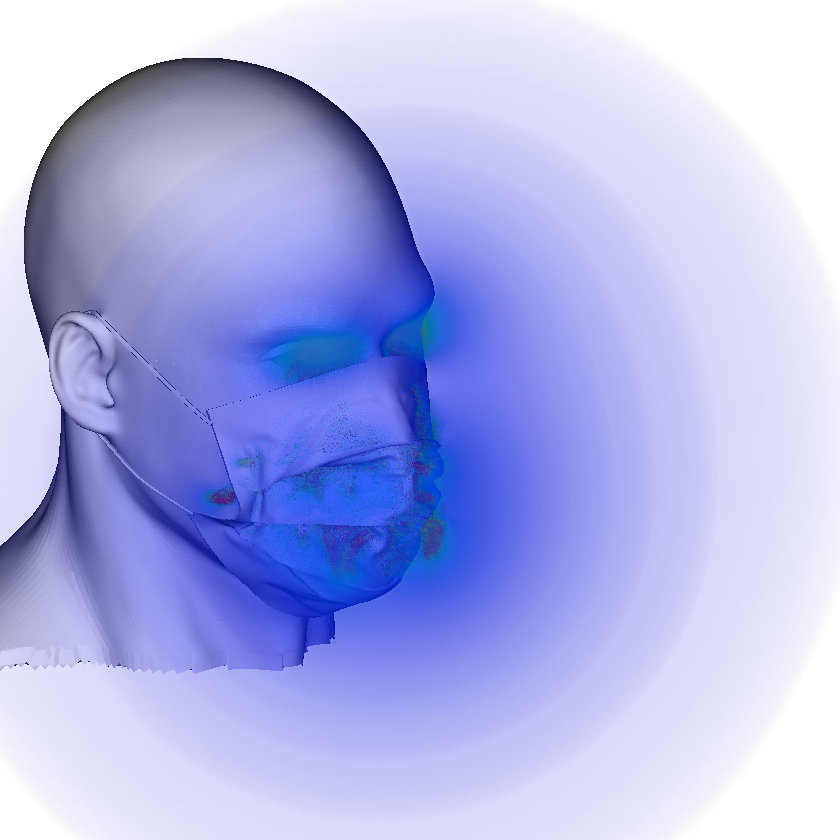}
    \subcaption{time = 0.05 s}
  \end{minipage}
  \begin{minipage}{0.3\hsize}
    \includegraphics[keepaspectratio,width=\textwidth]{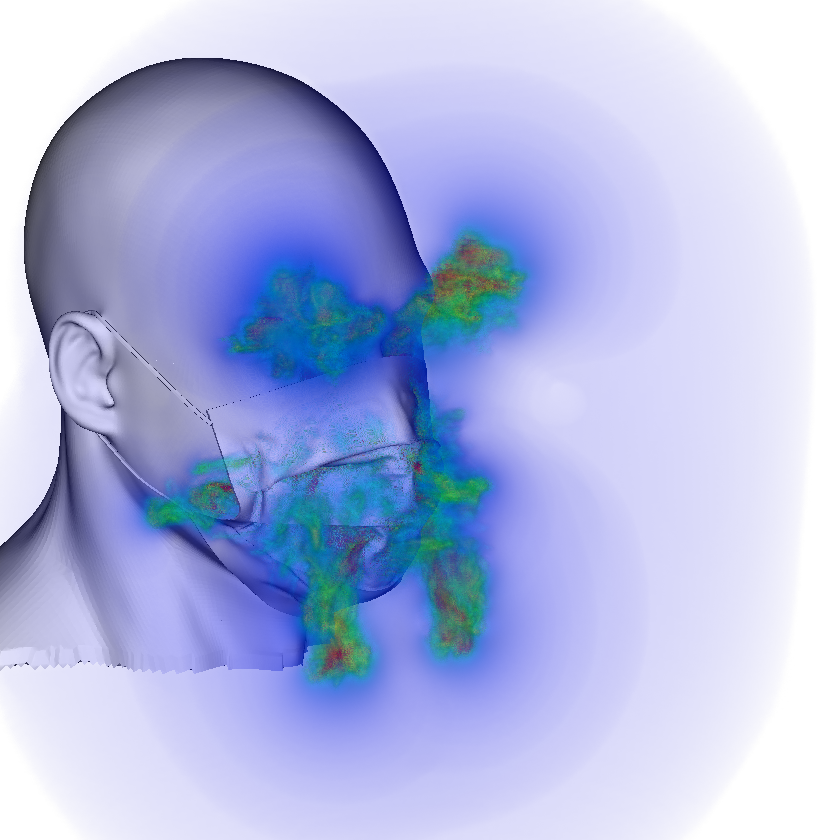}
    \subcaption{time = 0.13 s}
  \end{minipage}
  \begin{minipage}{0.3\hsize}
    \includegraphics[keepaspectratio,width=\textwidth]{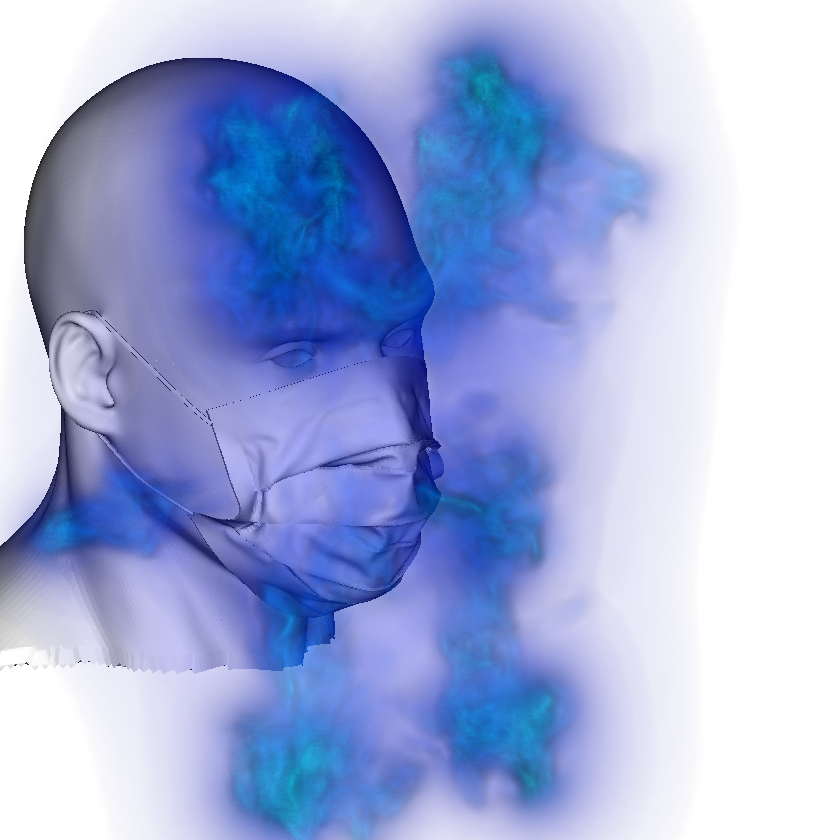}
    \subcaption{time = 0.65 s}
  \end{minipage}
  \caption{A volume-rendered 3D visualization colored by velocity magnitude.}
  \label{fig:surgical_volume}
\end{figure}

\begin{figure}[htb]
  \centering
  \begin{minipage}{0.3\hsize}
    \includegraphics[keepaspectratio,width=\textwidth]{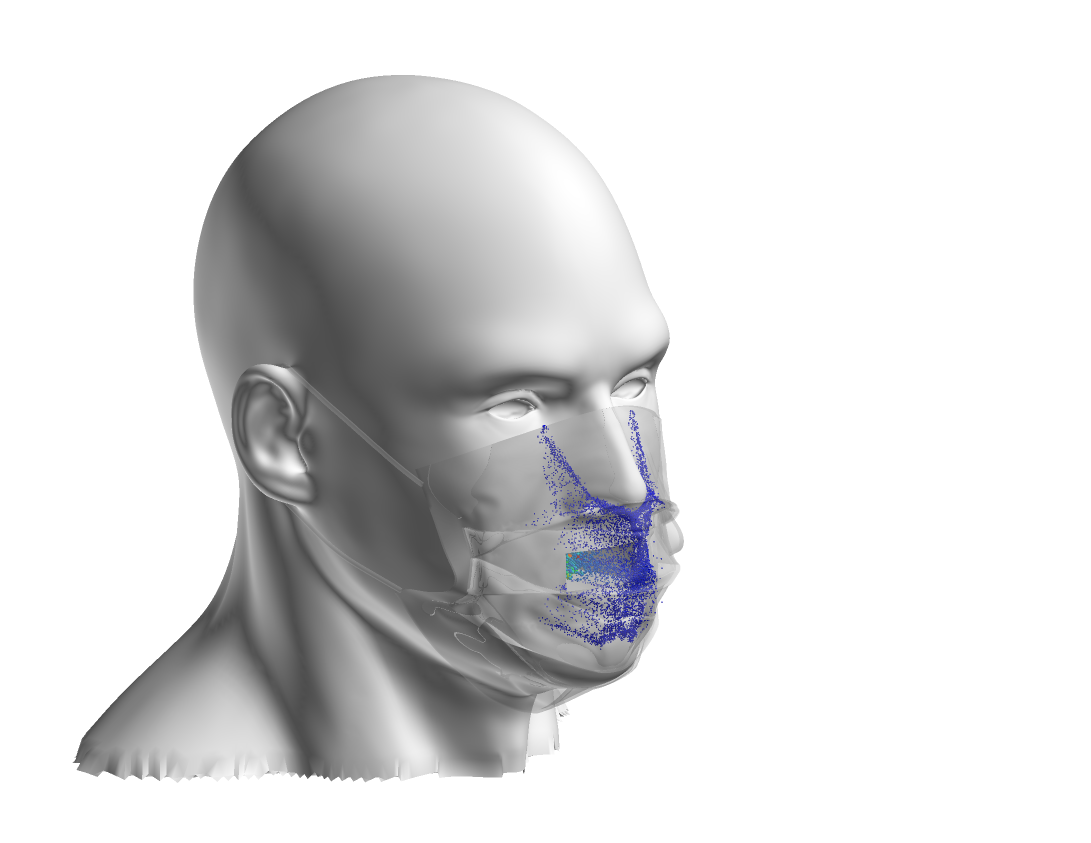}
    \subcaption{time = 0.05 s}
  \end{minipage}
  \begin{minipage}{0.3\hsize}
    \includegraphics[keepaspectratio,width=\textwidth]{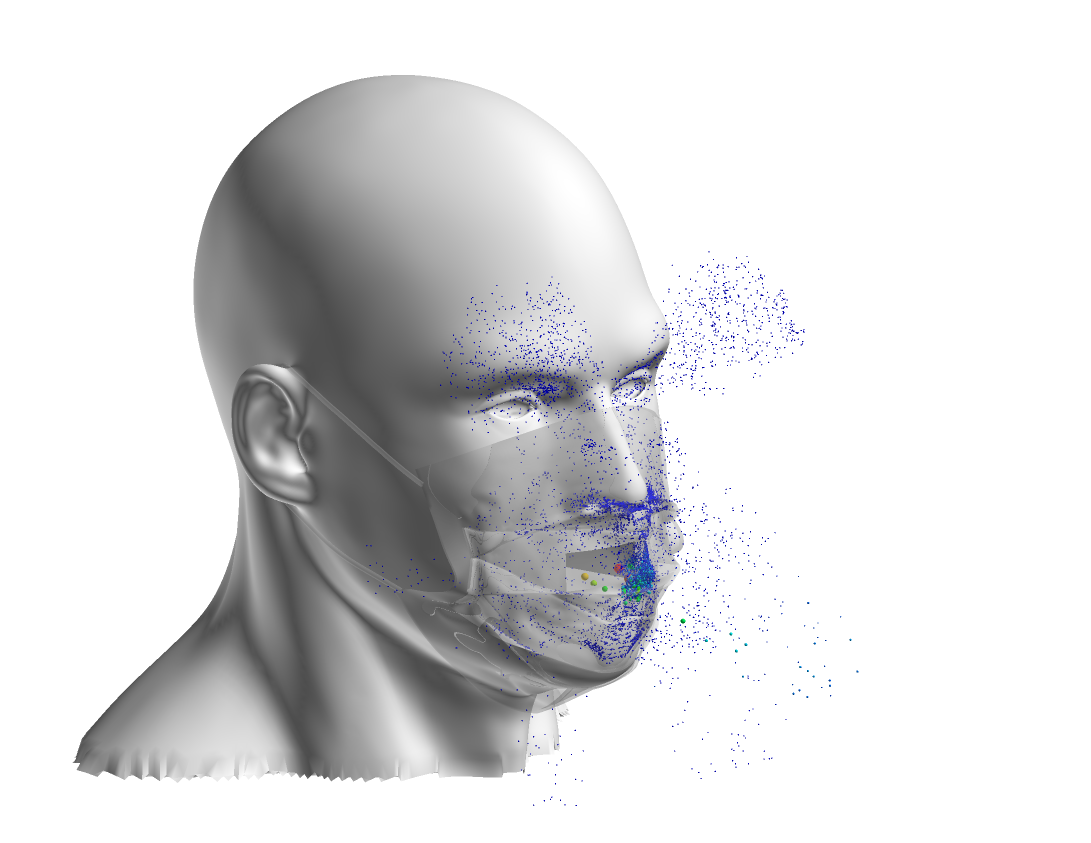}
    \subcaption{time = 0.13 s}
  \end{minipage}
  \begin{minipage}{0.3\hsize}
    \includegraphics[keepaspectratio,width=\textwidth]{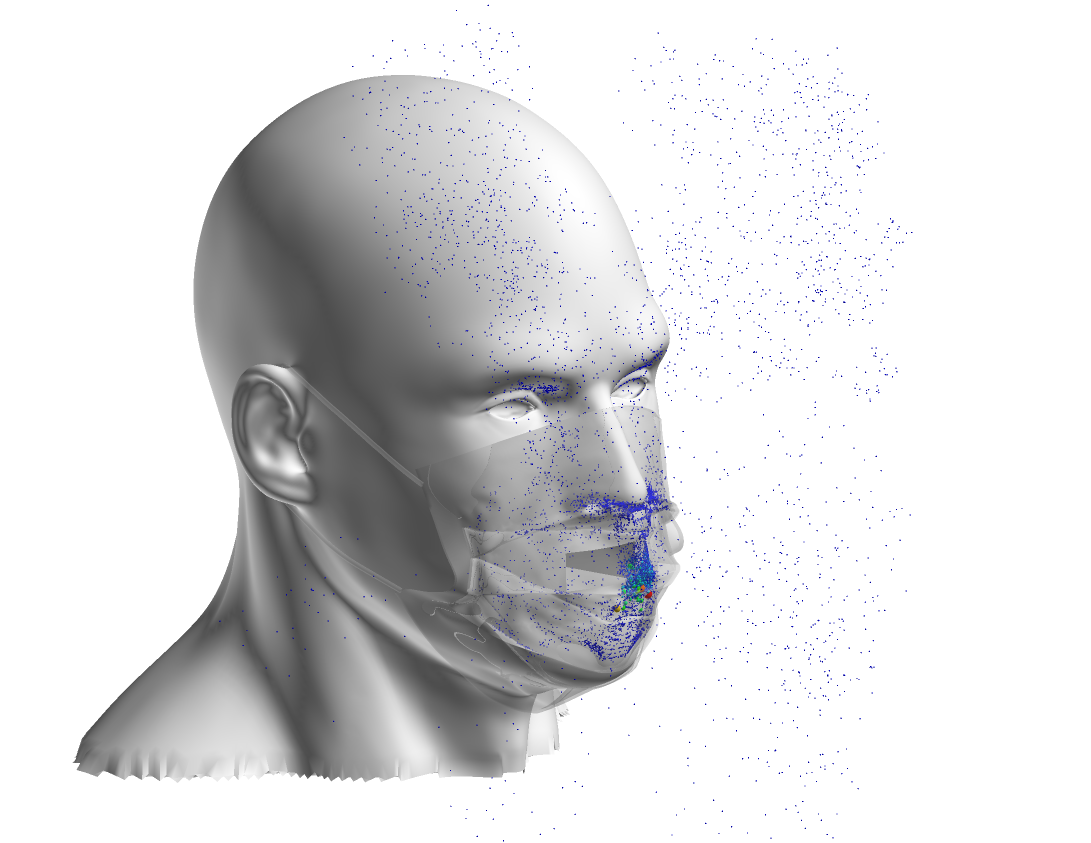}
    \subcaption{time = 0.65 s}
  \end{minipage}
  \caption{A spray particles distribution colored/sized by the droplet diameter.}
  \label{fig:surgical_spray}
\end{figure}

\begin{figure}[htb]
  \centering
  \begin{minipage}{0.3\hsize}
    \includegraphics[keepaspectratio,width=\textwidth]{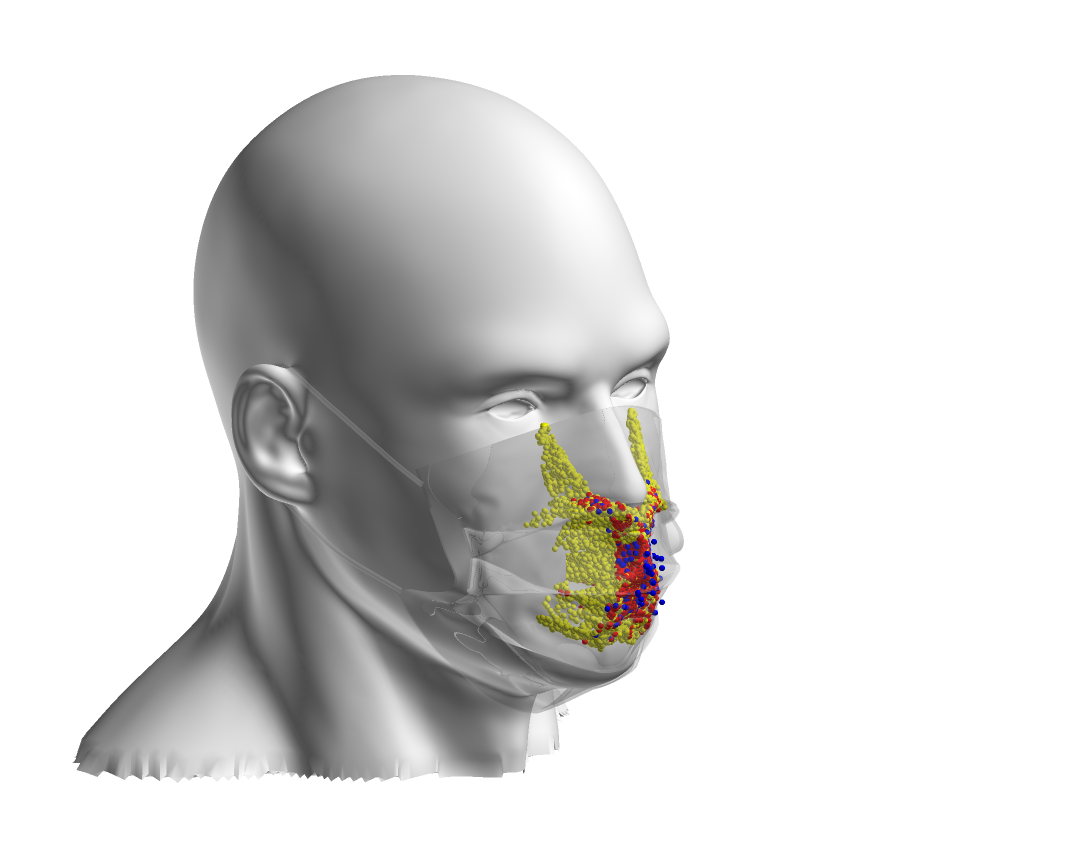}
    \subcaption{time = 0.05 s}
  \end{minipage}
  \begin{minipage}{0.3\hsize}
    \includegraphics[keepaspectratio,width=\textwidth]{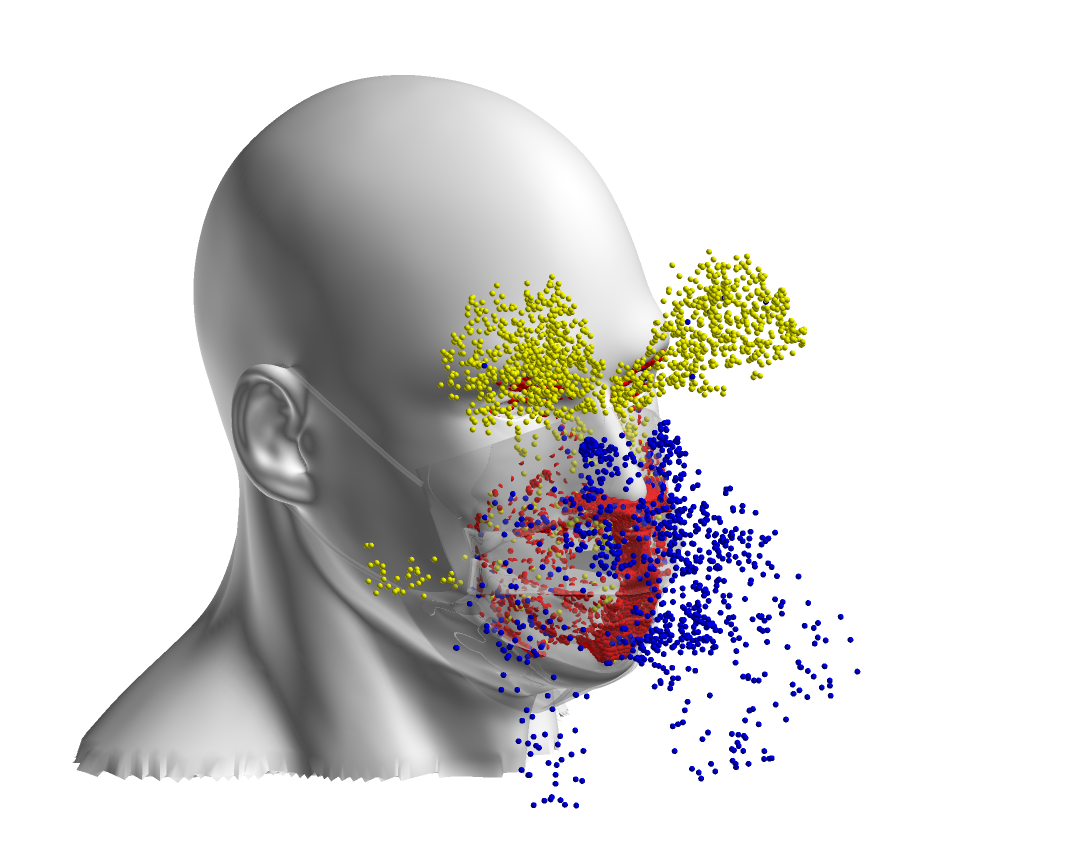}
    \subcaption{time = 0.13 s}
  \end{minipage}
  \begin{minipage}{0.3\hsize}
    \includegraphics[keepaspectratio,width=\textwidth]{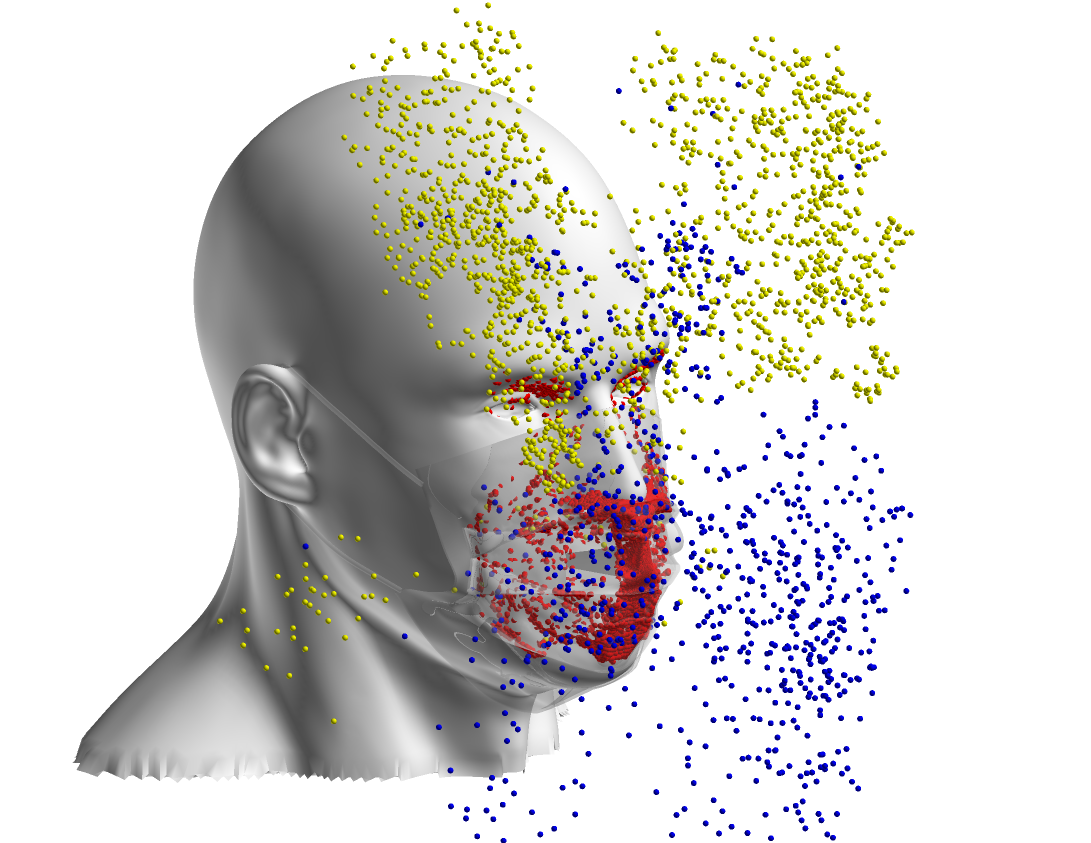}
    \subcaption{time = 0.65 s}
  \end{minipage}
  \caption{A spray particles distribution colored by the wall flag (yellow: fly, blue: penetrate, red: stick).}
  \label{fig:surgical_wallflag}
\end{figure}

 Figures \ref{fig:surgical_volume} and \ref{fig:surgical_spray} show the volume-rendered visualization and particle distribution of the 3D instantaneous (time = 0.05, 0.13 and 0.65 s) flow fields with a surgical mask. The two types of flow include a flow that leaks from the gap between the face and the mask surface, and a flow that permeates through the pressure loss from the entire mask surface. The gap between the face and mask forms a 3D complex shape, and these geometries are partially overlapped and narrow passed; however, the calculation method used in this study does not cause any problems. The initial particles placed in front of the mouth are diffused by the flow at every moment. These are separated into those captured by the mask surface (stick), penetrate the surface (penetrate), and leak through the gap (fly). Some particles are stuck on the face, and some flow to the region surrounded by the mouth and mask. Figure \ref{fig:surgical_wallflag} shows the particle distribution at a particle size normalized for clarity, and the color shows the state of the fly, penetrate, and stick. The collection rate of the particles captured by the face mask surface was obtained at the end of a single cough (time = 1.0 s).

\subsubsection{General surgical and cloth masks}

\begin{figure}[htb]
  \centering
  \begin{minipage}{0.4\hsize}
    \includegraphics[keepaspectratio,width=\textwidth]{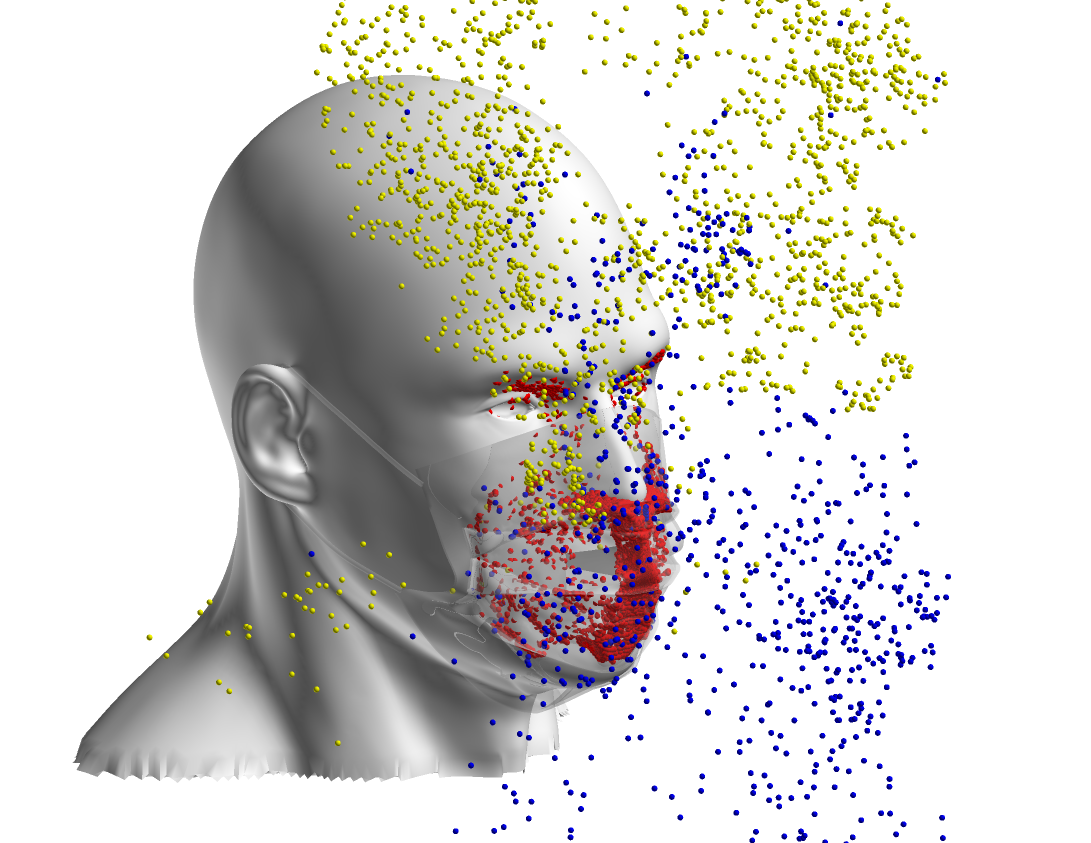}
    \subcaption{Surgical mask.}
    \label{fig:wallflag_surgical}
  \end{minipage}
  \hspace{0cm}
  \begin{minipage}{0.4\hsize}
    \includegraphics[keepaspectratio,width=\textwidth]{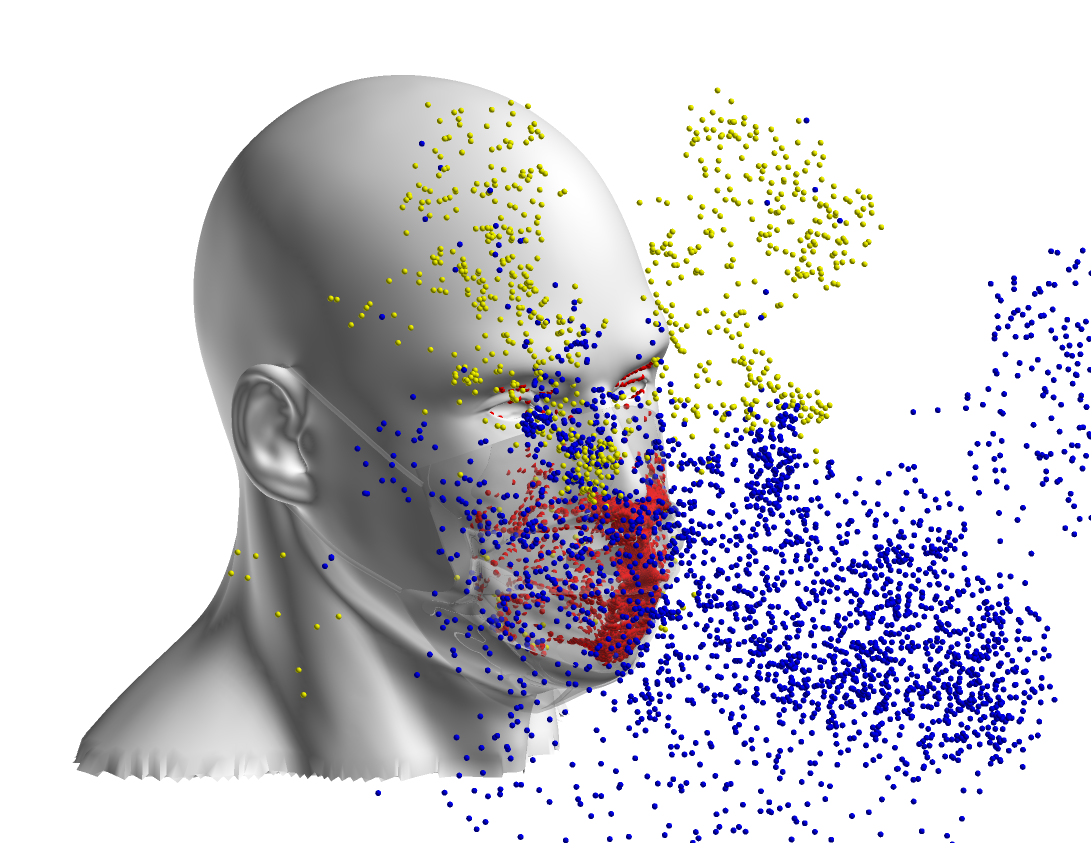}
    \subcaption{Cloth(polyester 100\%) mask.}
    \label{fig:wallflag_cloth}
  \end{minipage}
  \caption{
    Particle distribution after 1.0 s when wearing a surgical mask and fabric cloth mask, colored by flag (yellow: fly, blue: penetrate, red: stick).
  }
  \label{fig:wallflag_surgical_cloth}
\end{figure}

\begin{figure}[htb]
  \centering
  \begin{minipage}{0.4\hsize}
    \includegraphics[keepaspectratio,width=\textwidth]{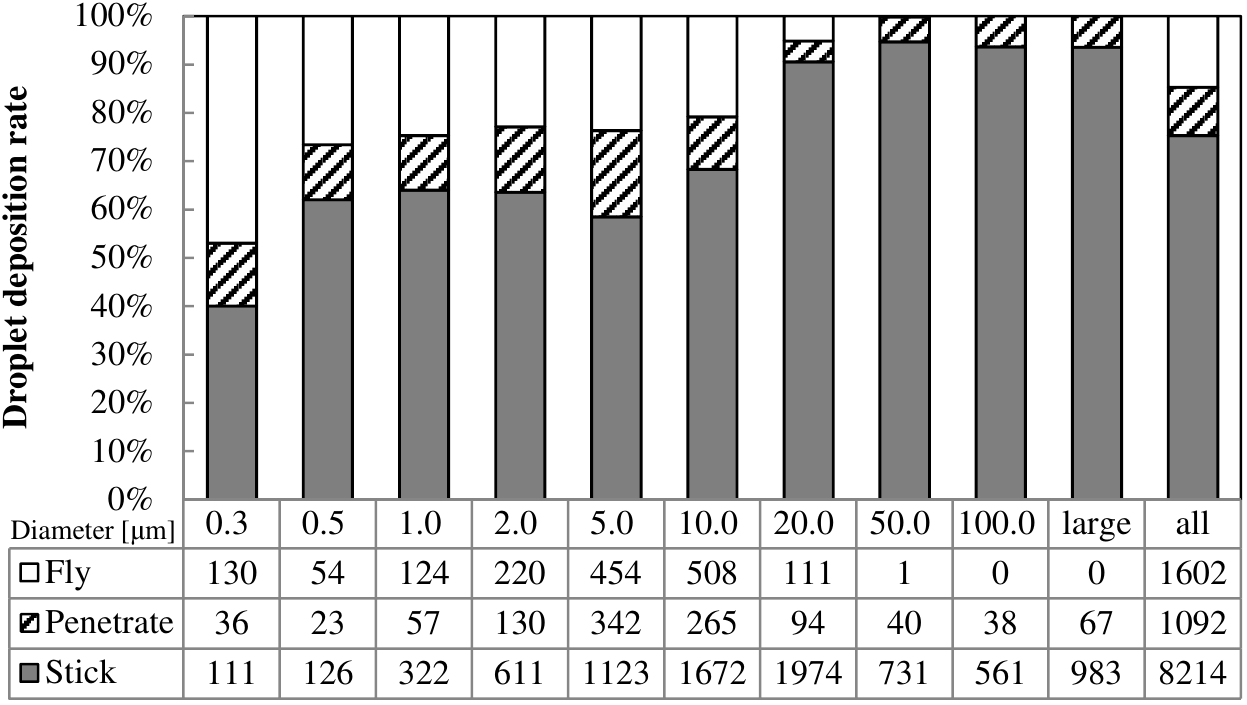}
    \subcaption{Surgical mask.}
    \label{fig:deposition_surgical}
  \end{minipage}
  \hspace{0cm}
  \begin{minipage}{0.4\hsize}
    \includegraphics[keepaspectratio,width=\textwidth]{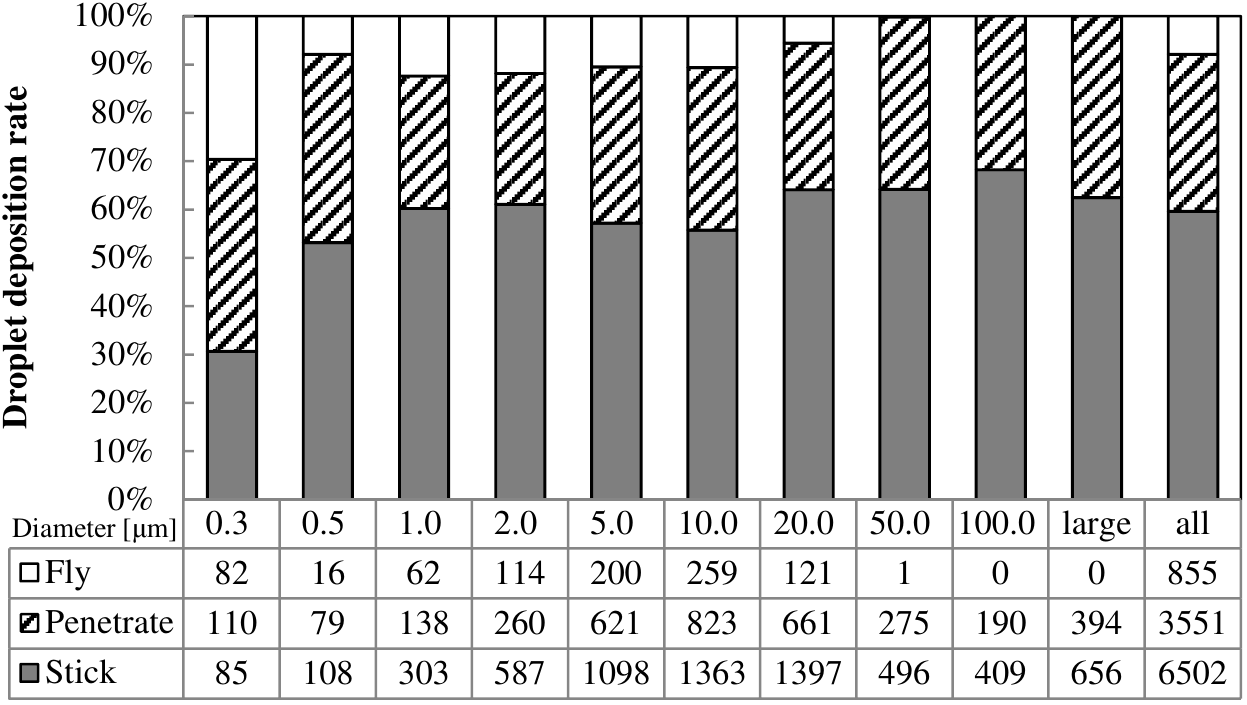}
    \subcaption{Cloth(polyester 100\%) mask.}
    \label{fig:deposition_cloth}
  \end{minipage}
  \caption{
    Particle deposition rate for each particle size.
  }
  \label{fig:deposition_surgical_cloth}
\end{figure}

 Figure \ref{fig:wallflag_surgical_cloth} shows the particle distribution after 1.0 s when wearing a surgical mask and fabric cloth mask. Figure \ref{fig:deposition_surgical_cloth} shows the results of the deposition rate counted with the wall flag information for each particle size. The cloth mask is a handmade mask made of 100\% polyester cloth, and its shape is the same as that of the surgical mask.
 In a general surgical mask, approximately 60-70\% of aerosol particles of 5 $\mu m$ or less are collected. Surgical masks exhibit a higher performance than that of cloth masks and fewer particles can pass through them. However, cloth masks have less leakage from the gap. Both mask types exhibited the same efficiency and approximately 70-80\% of the particles were collected in both mask types. In the evaluation by volume ratio, the contribution of large particles was dominant, resulting in a deposition rate of approximately 100\% for both mask types. To limit the measurement range within the fine particles ($< 20 \mu m$) of interest in the case of COVID-19, a volume deposition rate of approximately 80\% could be obtained, as could the particle number deposition rate.

\subsubsection{Differences in materials or type of masks}

\begin{figure}[htb]
  \centering
  \begin{minipage}{0.4\hsize}
    \includegraphics[keepaspectratio,width=\textwidth]{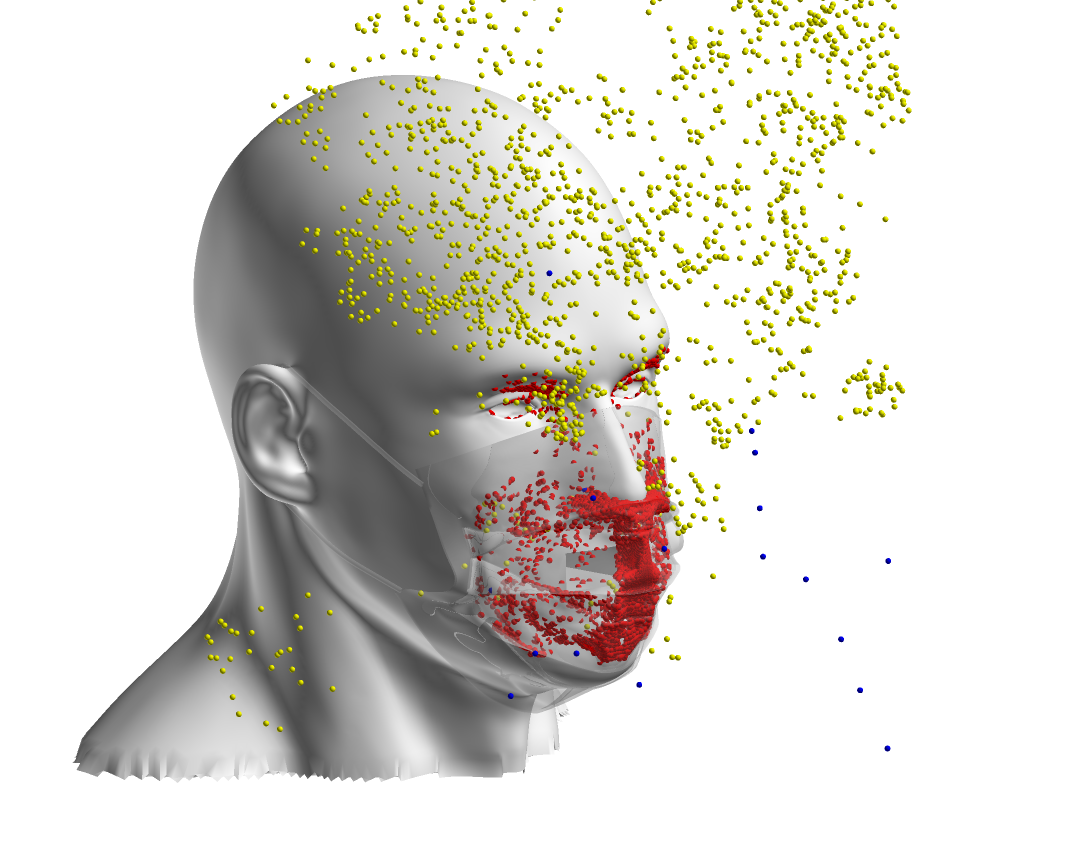}
    \subcaption{Surgical mask No.3, a high-performance type from different suppliers.}
    \label{fig:wallflag_surgical3}
  \end{minipage}
  \hspace{1.2cm}
  \begin{minipage}{0.4\hsize}
    \includegraphics[keepaspectratio,width=\textwidth]{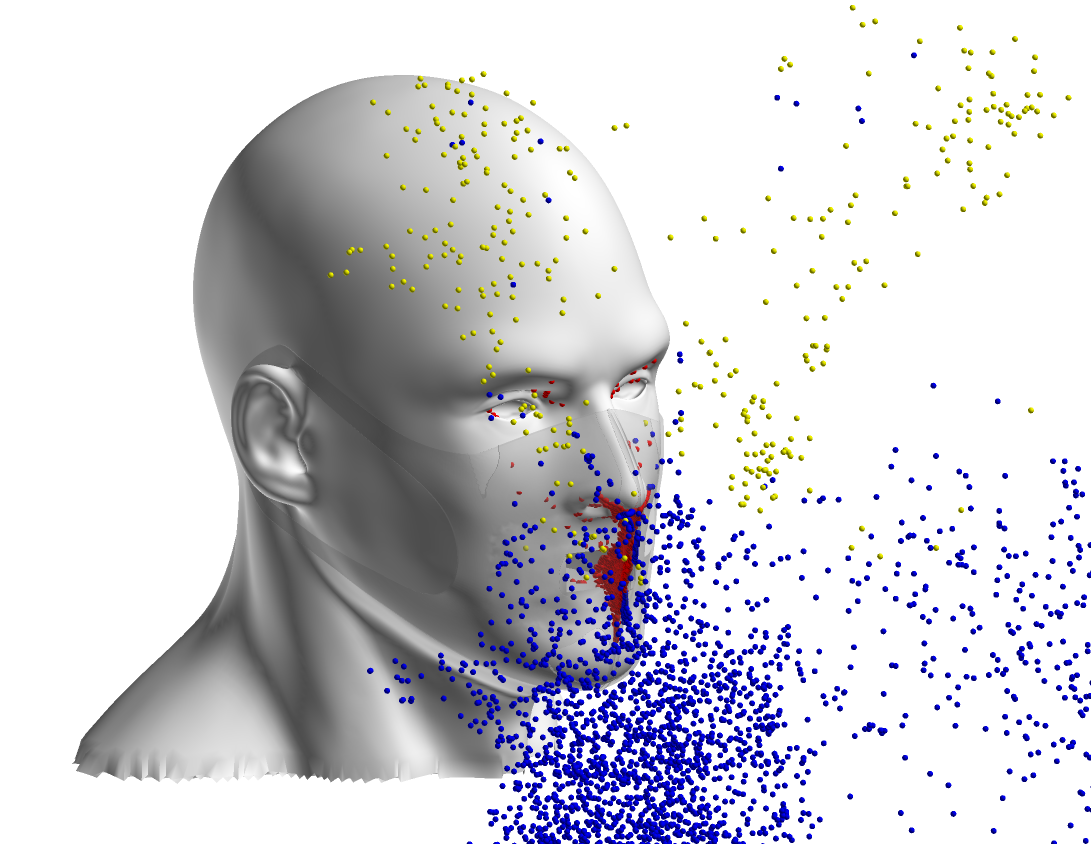}
    \subcaption{Urethane mask made from poly-urethane material.}
    \label{fig:wallflag_urethane}
  \end{minipage}
  \hspace{0cm}
  \begin{minipage}{0.4\hsize}
    \includegraphics[keepaspectratio,width=\textwidth]{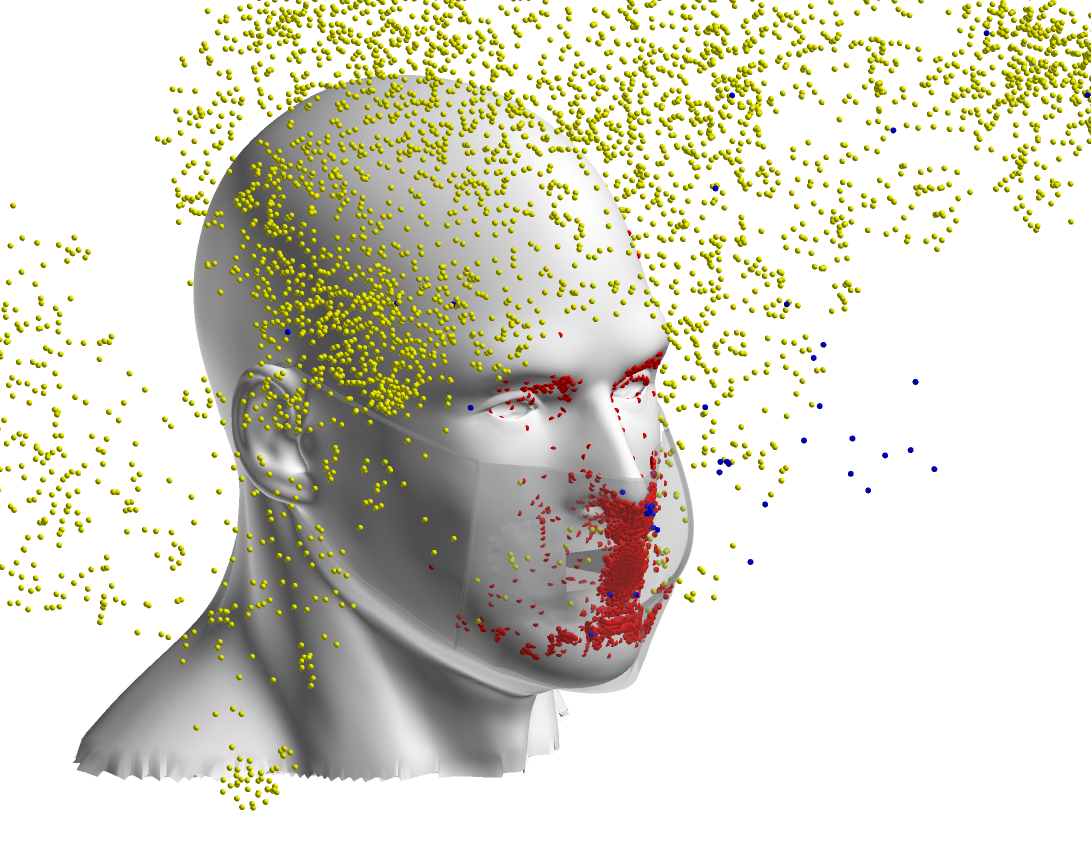}
    \subcaption{Gauze mask.}
    \label{fig:wallflag_gauze}
  \end{minipage}
  \hspace{0cm}
  \begin{minipage}{0.4\hsize}
    \includegraphics[keepaspectratio,width=\textwidth]{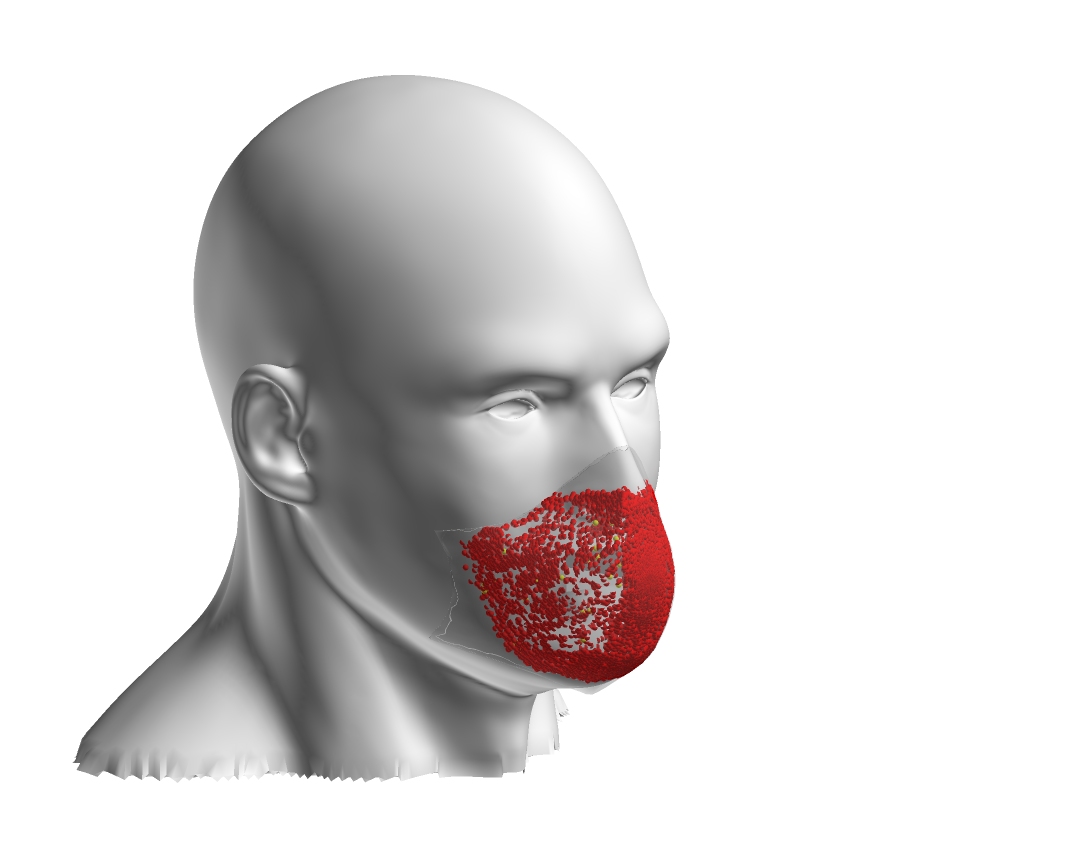}
    \subcaption{N95 mask.}
    \label{fig:wallflag_N95}
  \end{minipage}
  \caption{
    Particle distribution after 1.0 s for various types of face masks, colored by flag (yellow: fly, blue: penetrate, red: stick).
  }
  \label{fig:wallflag_variety}
\end{figure}

 Figure \ref{fig:wallflag_variety} shows the particle distribution after 1.0 s for the four types of face masks. Surgical mask No.3 is another surgical mask product supplied by another manufacturer, which is labeled as a high-performance type. The urethane mask is a polyurethane mask designed for easy breathing. The gauze mask is a mask distributed by the Japanese government to the general public, created by folding several layers of gauze filter media. The N95 mask is a type of mask that is used in the medical field and is in perfect contact with the face.

\begin{figure}[htb]
  \centering
  \begin{minipage}{0.7\hsize}
      \includegraphics[keepaspectratio,width=\textwidth]{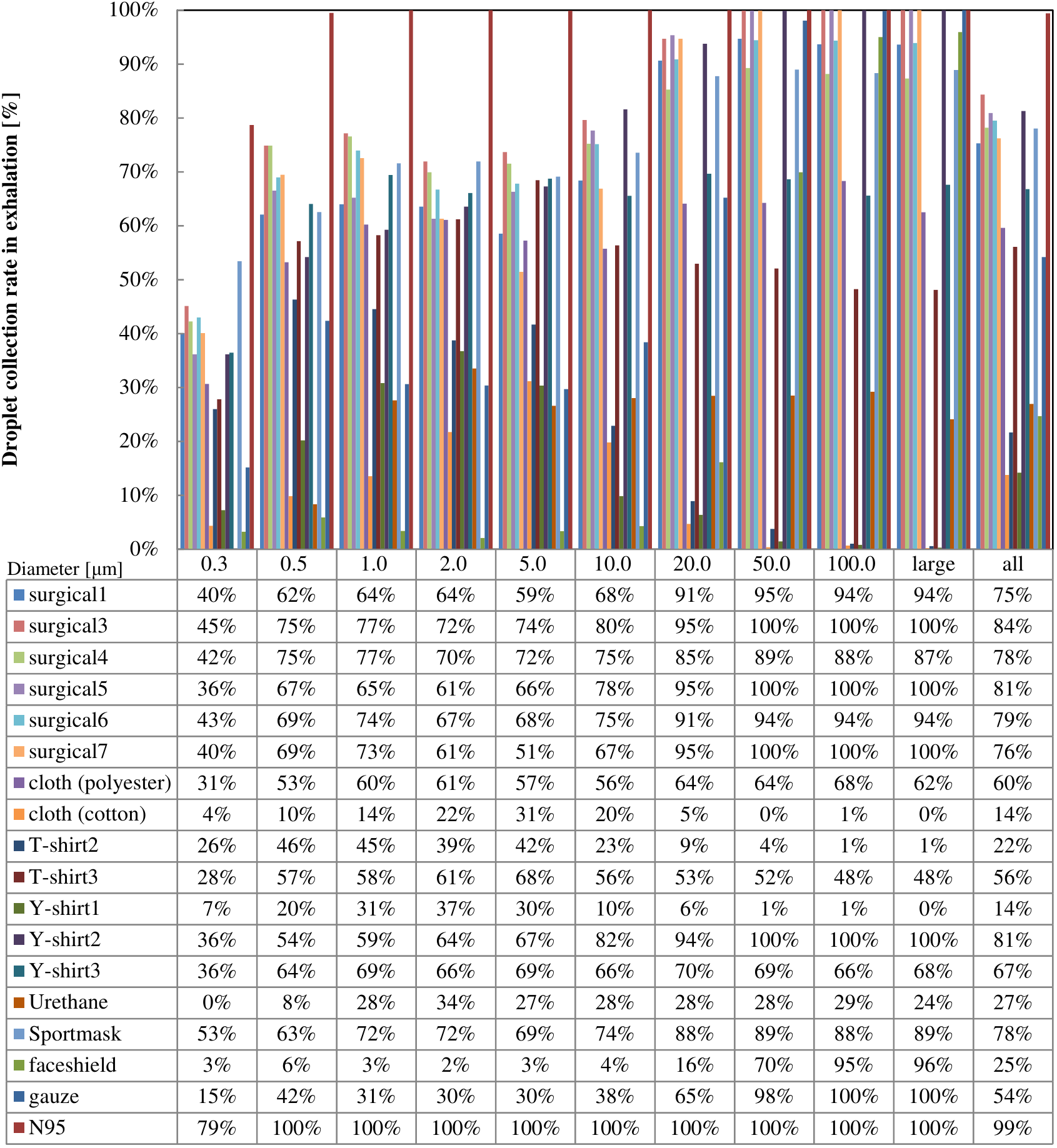}
  \end{minipage}
  \caption{
     A comparison between the collection rates for each material/mask type in the exhalation event.
  }
  \label{fig:eff_material}
\end{figure}

 Figure \ref{fig:eff_material} compares the collection rates (calculated from the deposition rate) of various mask types. The surgical masks of No.3-7 are commercially available masks manufactured by different manufacturers with three layers of polyolefin nonwoven fabric, and cloth masks are handmade masks composed of two layers of 100\% polyester cloth. T-shirt masks are handmade masks made from one to three layers of used cotton T-shirts, and Y-shirt masks are handmade masks that use one to three layers of cotton/polyester-blended shirt fabrics. In addition, the shapes of these mask types are the same in the calculation.

 The surgical masks vary depending on the product; however, all products exhibit an overall collection rate of approximately 80\%. The collection rate of aerosol particles of 5 $\mu m$ or less is approximately 40-70\%, which is far less than that of the filter itself (99\% cut, etc.). This indicates that the aerosol particles are easily carried by the flow and leak mainly from the gap without passing through the filter surface. Therefore, when considering the overall performance of masks, the performance of the filter and the prevention of particle leakage from the gaps must be considered. The particle images show that leakage from the gap beside the nose became dominant depending on how the mask was worn. We used a surgical mask that was made in Japan. It generally has a shaping device that closes the gaps; thus, a similar performance was expected to be obtained; nonetheless, a similar performance cannot be anticipated when the mask is not installed properly or the mask is stored in an unacceptable condition.

 The T-shirt masks that are made of cotton fabric of used T-shirts, the highest achievable collection rate of aerosol particles is only approximately 30\% when there is only one layer, and the overall performance is reduced to approximately one-third of that of the polyester cloth mask. The multi-layered structure improves the collection rate performance, and the 3-layered structure provides approximately the same performance as the 2-layered polyester cloth mask. Similarly, the highest achievable collection rate of aerosol particles in a Y-shirt mask is only approximately 37\% when only one layer of cloth is used, and the performance can be improved by using two layers. However, the collection rate of large particles decreased in the case of 3-layered Y-shirt, and the overall collection rate decreased by approximately 14\%. This is because the pressure loss became extremely large, and the number of large particles leaking from the gap increased. The balance between the pressure loss and filter collection rate must be considered in the evaluation of the mask performance.

 Additionally, when the transmissive conditions presented by Sergey et al. \cite{Sergey2009} and Aydin et al. \cite{Aydin2020} were used, the overall collection rate of the surgical and handmade cloth masks was 80 and 82\%, respectively. The reason is because distribution data for each particle size was not presented \cite{Aydin2020}, only the representative values are used, and the mask shape and pressure loss are assumed to be identical. Thus, the performance is slightly overestimated. This is because insufficient data, such as a collection rate of $1 \mu m$ or more, were created by interpolation, and the collection rate for the aerosol reacted sensitively. In this analysis, the accuracy of the collection rate data for each aerosol particle size must be known.

\begin{figure}[htb]
  \centering
  \begin{minipage}{0.7\hsize}
      \includegraphics[keepaspectratio,width=\textwidth]{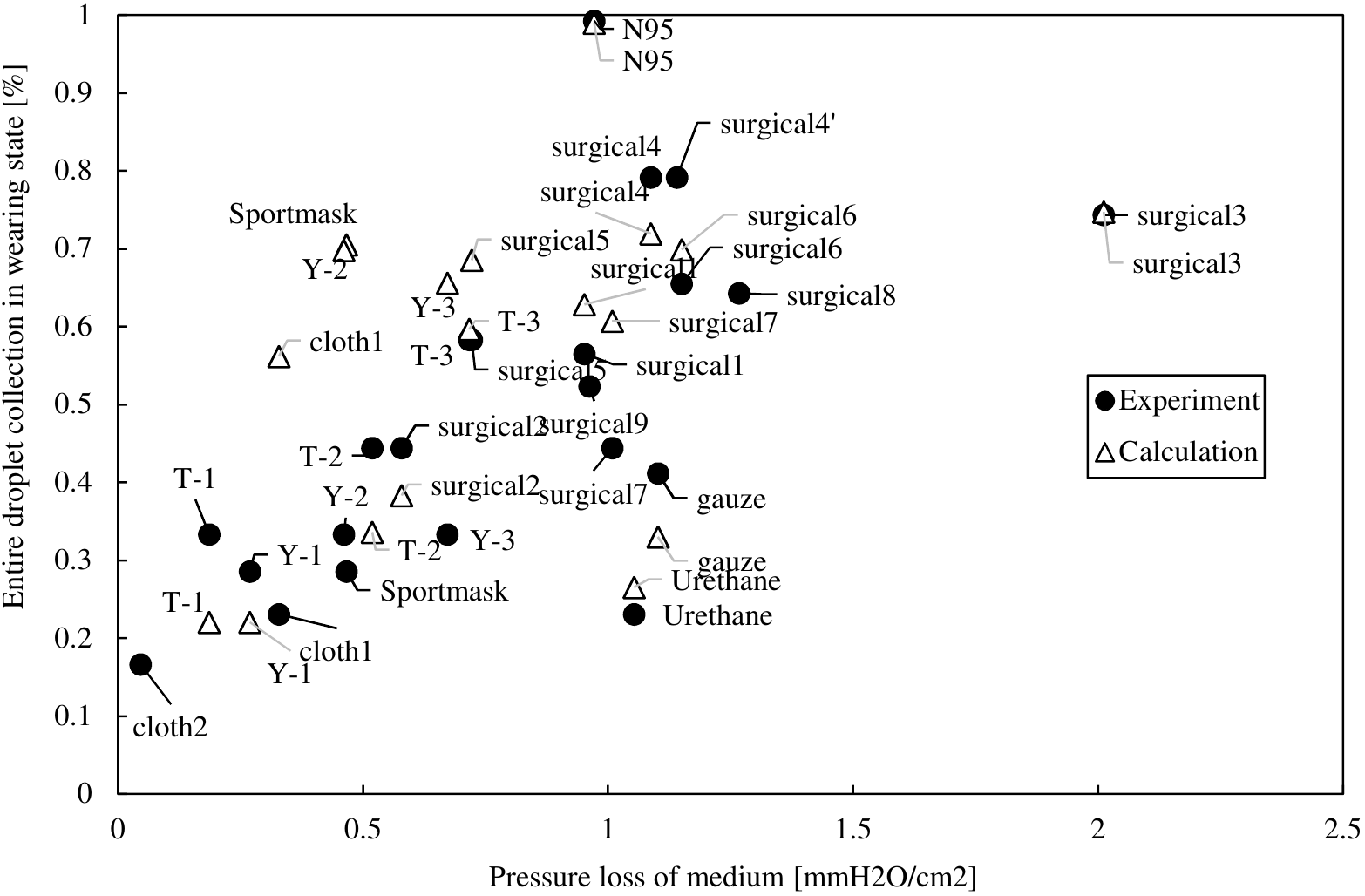}
  \end{minipage}
  \caption{
     Comparison of the collection rates with the experimental results.
  }
  \label{fig:exp_comparison}
\end{figure}

 Figure \ref{fig:exp_comparison} demonstrates a comparison between the entire mask performance with our experimental results for various types of masks. The performance was measured within less than $10 \mu m$ diameter because of the limitations imposed by the experimental measurement equipment. In addition, the measured performance are summarized by the filter breathability (pressure loss) of the medium. Surgical masks have a high detection performance; however, they form a group with a large pressure loss and difficulty in breathing. Cloth masks allow air to easily pass through, while collection efficiency is low and shows a strong correlation between the breathability and mask performance. The urethane masks are superior in term of breathability; however, these are clearly less efficient than other masks.

 The calculations reveal qualitatively reasonable trends; however, several mask types have been overestimated such as the Y-shirt2 and cloth1 cases. This is because the phenomenon in which the droplets are split by the fibers, leading to an increase in the number from the initial value was observed during the experiments; thus, the phenomenon was not considered in this study. Consequently, in future investigations, we must consider a model that takes into account droplet splitting under such conditions.

\subsubsection{Worst and best cases}\label{worst_and_best}

\begin{figure}[htb]
  \centering
  \begin{minipage}{0.4\hsize}
    \includegraphics[keepaspectratio,width=\textwidth]{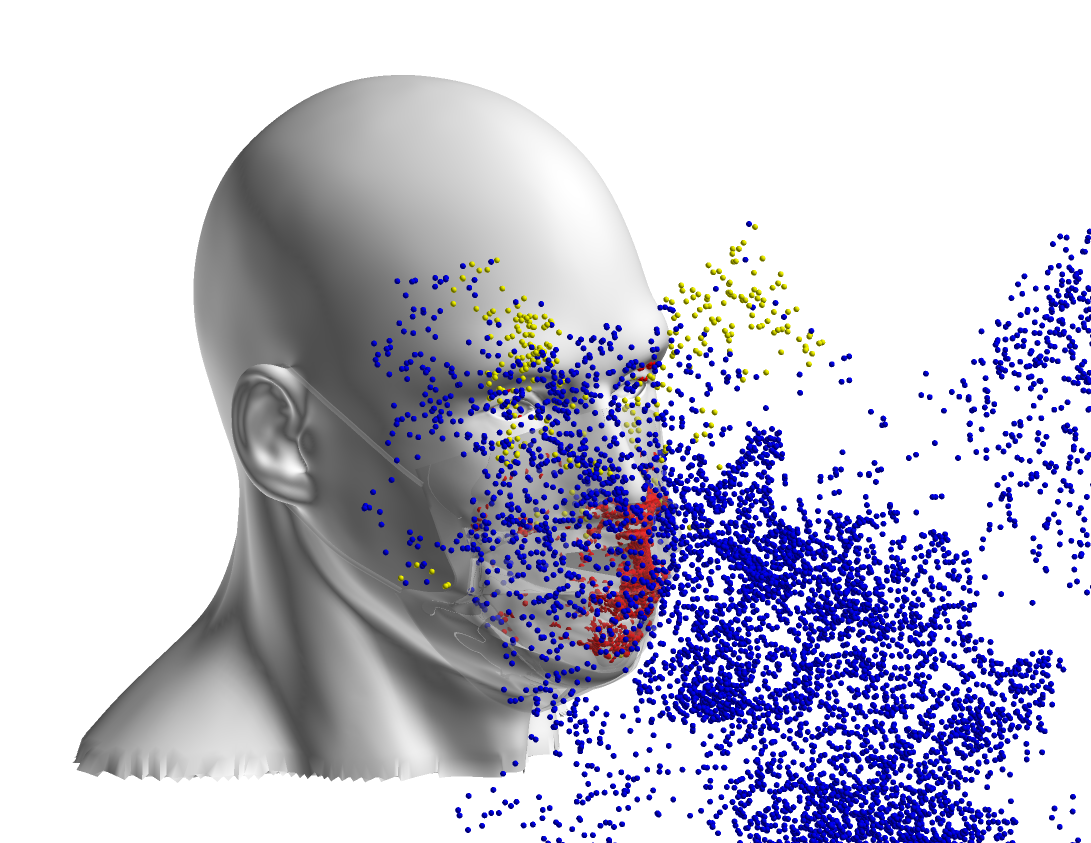}
    \subcaption{Cloth(cotton 100\%) mask.}
    \label{fig:wallflag_T-shirt1}
  \end{minipage}
  \hspace{0cm}
  \begin{minipage}{0.4\hsize}
    \includegraphics[keepaspectratio,width=\textwidth]{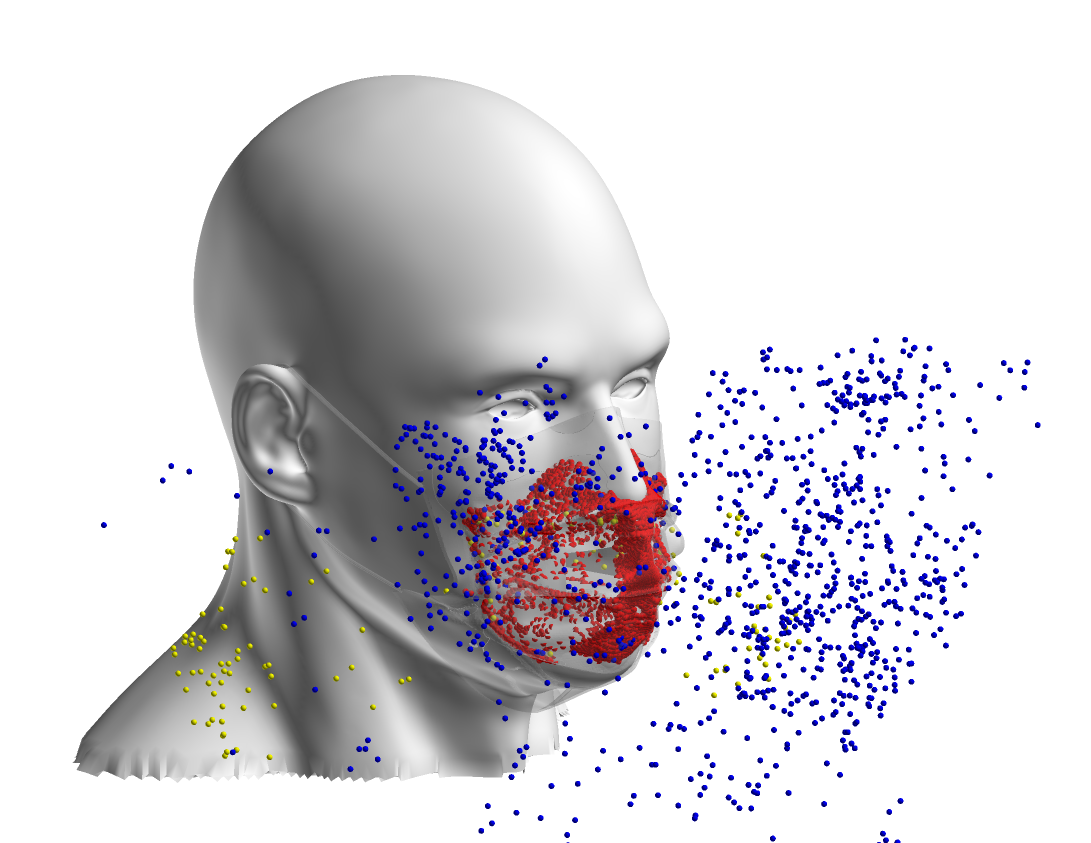}
    \subcaption{Surgical mask wearing in the fit state.}
    \label{fig:wallflag_surgical_fit}
  \end{minipage}
  \caption{
    Particle distribution after 1.0 s when wearing a cloth(cotton) mask and a surgical mask in the fit state, colored by flag (yellow: fly, blue: penetrate, red: stick).
  }
  \label{fig:wallflag_T-shirt1_surgical_fit}
\end{figure}

\begin{figure}[htb]
  \centering
  \begin{minipage}{0.4\hsize}
    \includegraphics[keepaspectratio,width=\textwidth]{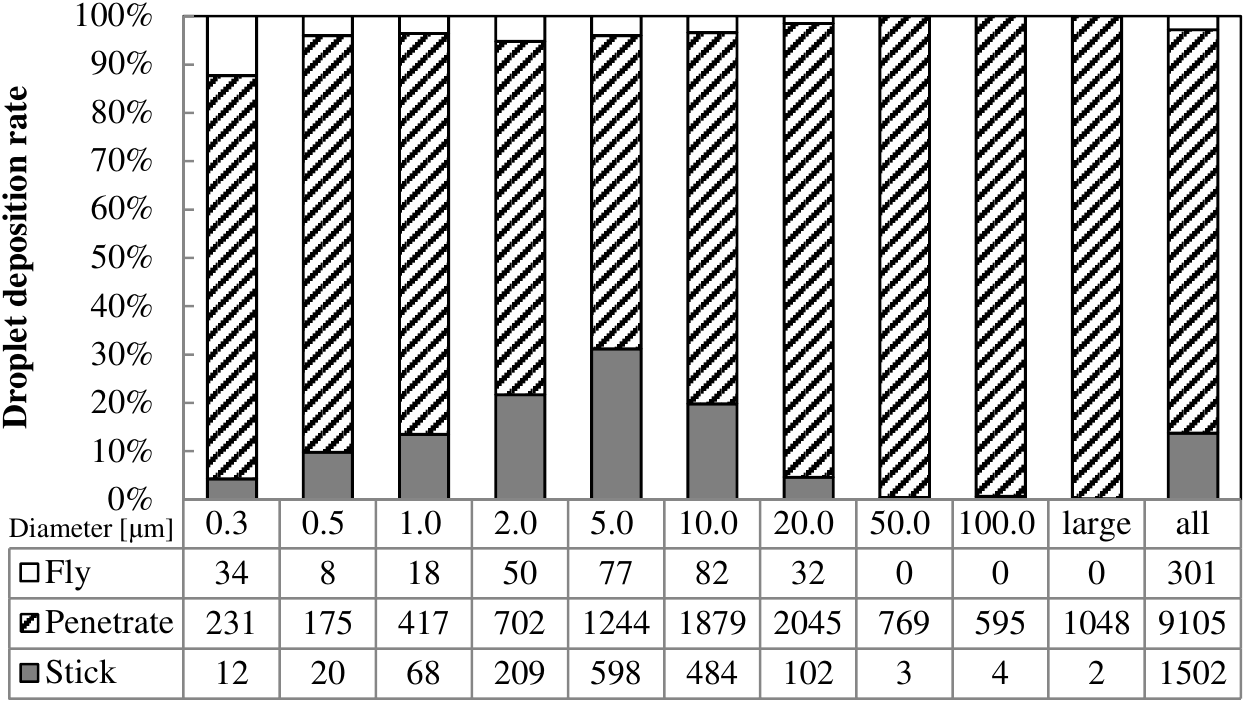}
    \subcaption{Cloth(cotton 100\%) mask.}
    \label{fig:depositionT-shirt1}
  \end{minipage}
  \hspace{0cm}
  \begin{minipage}{0.4\hsize}
    \includegraphics[keepaspectratio,width=\textwidth]{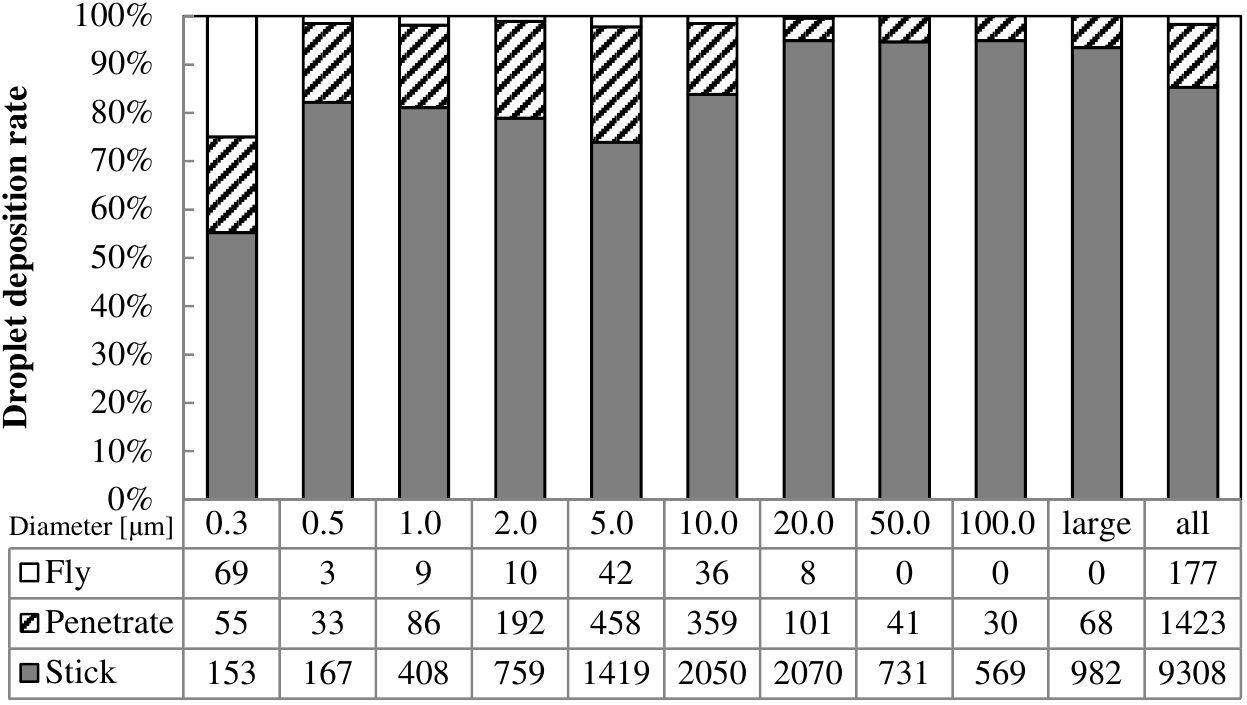}
    \subcaption{Surgical mask wearing in the fit state.}
    \label{fig:deposition_surgical_fit}
  \end{minipage}
  \caption{
    The particle deposition rate for each particle size.
  }
  \label{fig:deposition_T-shirt1_surgical_fit}
\end{figure}

 The worst case was achieved by the cloth mask made of cotton, which performed equivalent to 1-layered T-shirt. In the best case, the surgical mask shape was set to fit the face, which close the gap between the nose and mask and maintains the pressure loss and transmission rate the same as those of the conventional surgical mask. The results of droplet distribution after 1.0 s are shown in Fig. \ref{fig:wallflag_T-shirt1_surgical_fit}. Figure \ref{fig:deposition_T-shirt1_surgical_fit} illustrates the results of the deposition rate for each particle size.
In the worst case, average deposition rate of aerosol particles was only approximately 16\%. In the best case, the number of leaking particles was small, and most of them were collected by the mask, which was equivalent to the performance of the filter medium. This illuminates the importance of wearing a mask without any gaps between the face and mask.

\subsubsection{Face shield}

\begin{figure}[htb]
  \centering
  \begin{minipage}{0.4\hsize}
    \includegraphics[keepaspectratio,width=\textwidth]{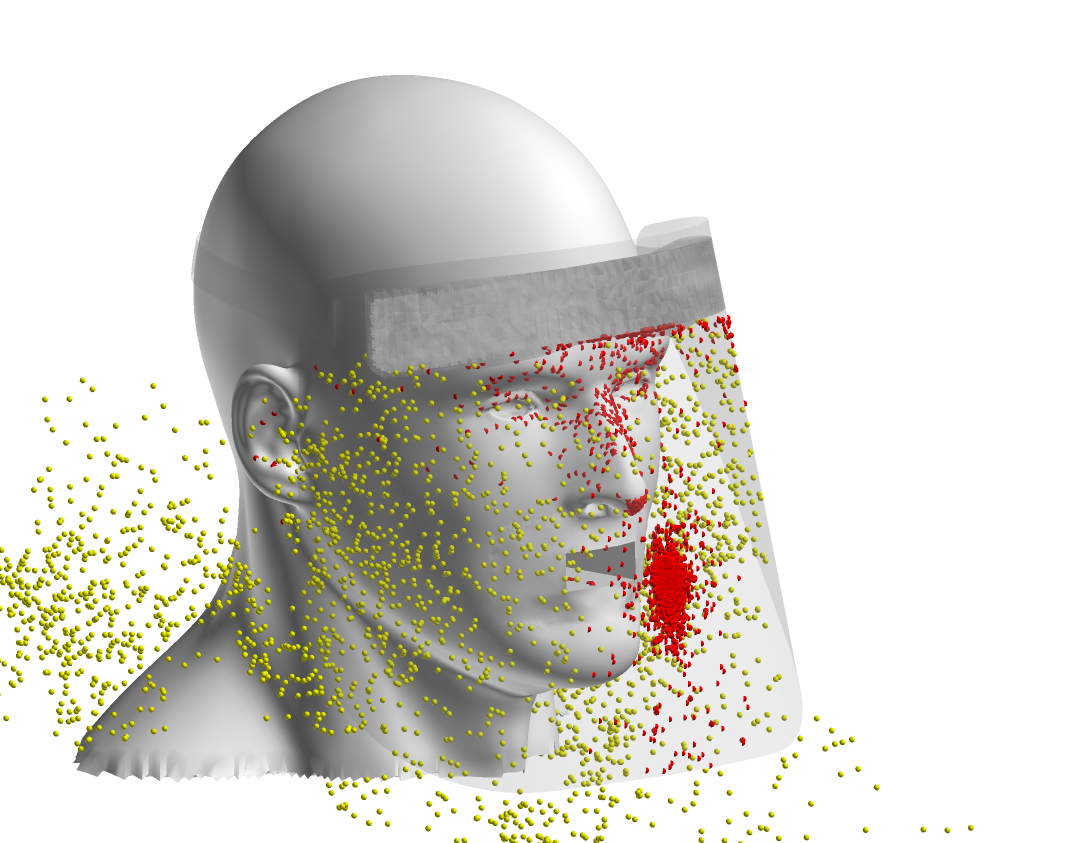}
    \subcaption{Particle distribution after 3.0 s when wearing a face shield.}
    \label{fig:wallflag_faceshield}
  \end{minipage}
  \hspace{0cm}
  \begin{minipage}{0.4\hsize}
    \includegraphics[keepaspectratio,width=\textwidth]{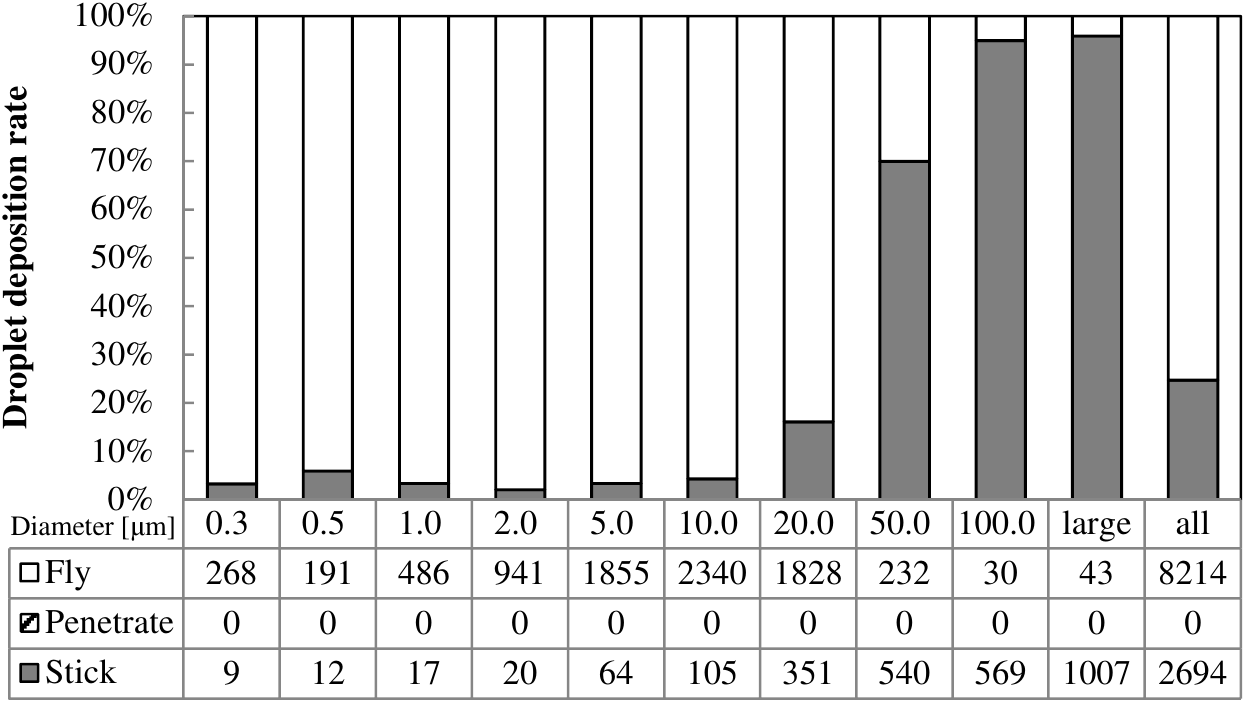}
    \subcaption{Particle deposition rate when wearing a face shield.}
    \label{fig:deposition_faceshield}
  \end{minipage}
  \caption{
    Particle distribution and deposition in the case of face shield.
  }
  \label{fig:faceshield}
\end{figure}

 A model with a face shield attached to the head was used as an alternative to the mask. The transmission model was removed, and only the particles that were adsorbed on the wall surface by inertial motion were evaluated.
 Figure \ref{fig:faceshield} shows the particle distribution after $3.0 s$ and the result of the collection rate for each particle size for the face shield. Approximately 80\% of the particles leak, and most aerosol particles with a diameter of $5 \mu m$ or less leak. This indicates that the face shield cannot replace the mask in terms of exhalation protection. Moreover, the flying particles include those that remain within the face shield.

\subsubsection{Double mask}

\begin{figure}[htb]
  \centering
  \begin{minipage}{0.7\hsize}
      \includegraphics[keepaspectratio,width=\textwidth]{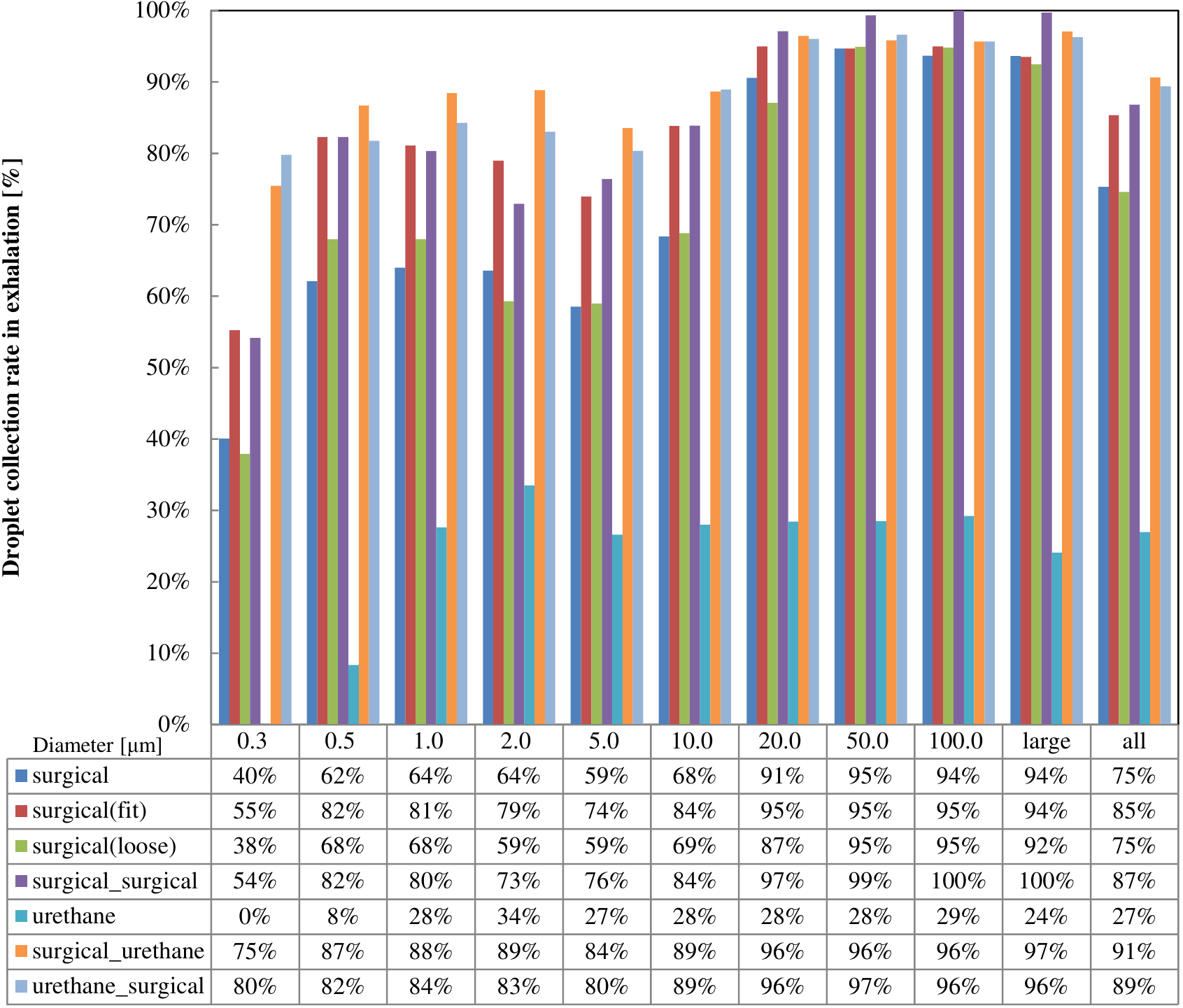}
  \end{minipage}
  \caption{
     Comparison of the collection rates for various droplet sizes in double-mask configurations for exhalation protection.
  }
  \label{fig:eff_double_exhale}
\end{figure}

 The performance of was evaluated using a double mask. Figure \ref{fig:eff_double_exhale} shows the collection rate for each droplet size in double-mask configurations; for instance, \textit{surgical\_urethane} indicates that the surgical and urethane masks are double-mounted, where the urethane mask is mounted over the inner surgical mask. The shape of a double mask is obtained using the structural analysis when the double mask is worn in the same manner as described in Subsection \ref{numerical_method_exhalation}.
 In the case of double mask, with two surgical masks or mixed usage of surgical and urethane masks, an improvement of approximately 10 to 15\% in the particle collection performance was observed, which was owing to the narrowing the gaps between the faces by pressing with the outer mask. The collection rate was approximately equivalent to the single mask in close contact with the face, as described in Subsection \ref{worst_and_best}. Therefore, suppressing the gap between the face and masks is important.

\subsection{Inhalation block efficiency evaluation}

\begin{figure}[htb]
  \centering
  \begin{minipage}{0.4\hsize}
      \includegraphics[keepaspectratio,width=\textwidth]{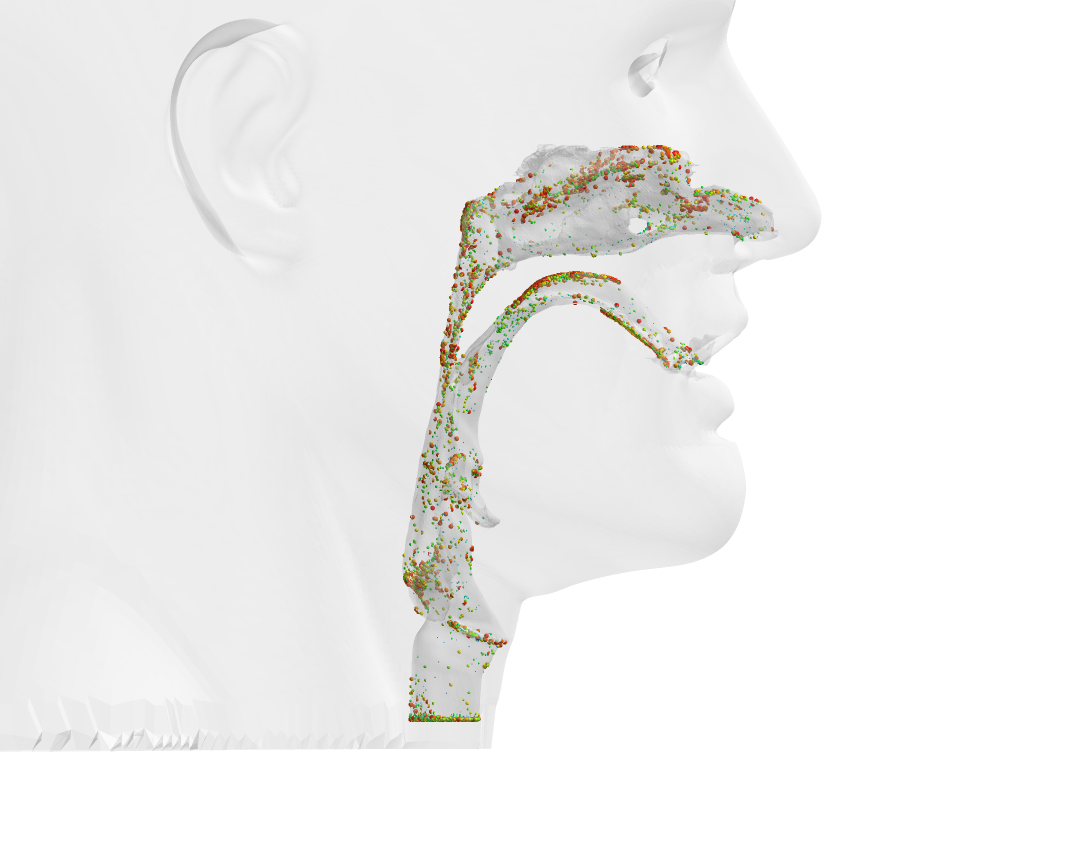}
  \end{minipage}
  \hspace{0cm}
  \begin{minipage}{0.4\hsize}
      \includegraphics[keepaspectratio,width=\textwidth]{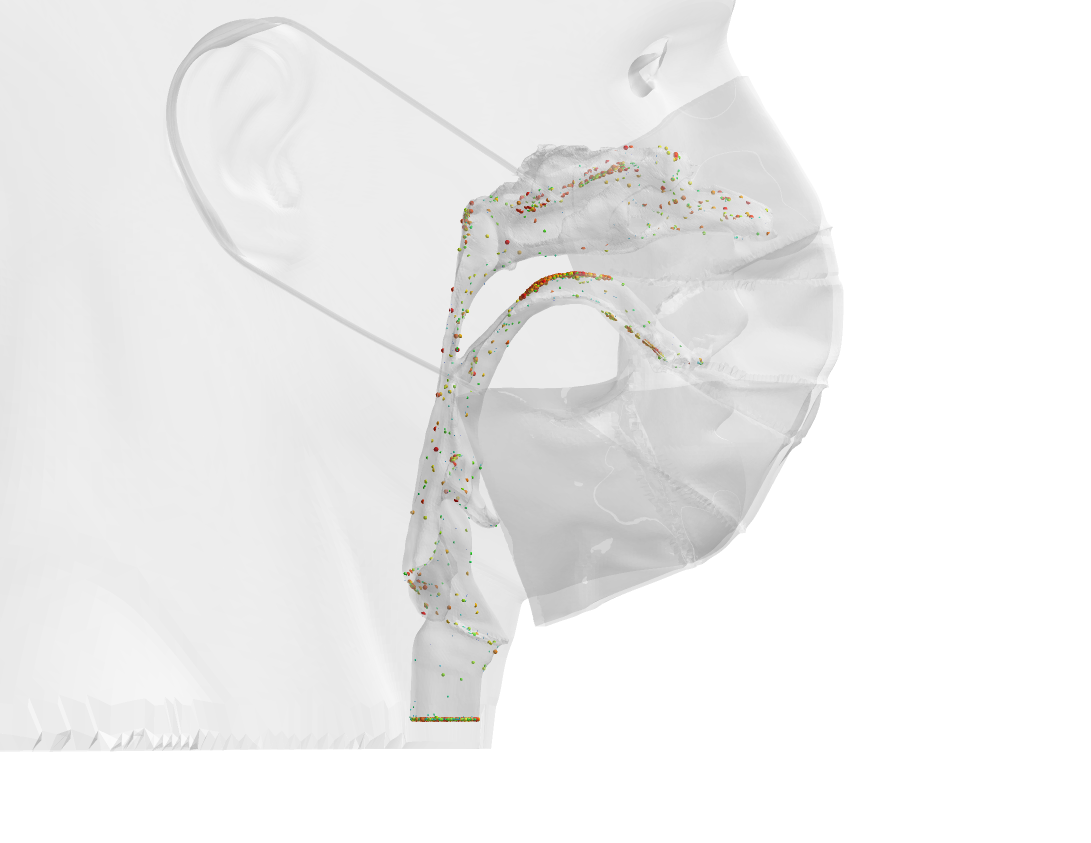}
  \end{minipage}
  \caption{
     Droplet distribution after 6.0s colored by diameter; left: no mask, right: surgical mask(fit).
  }
  \label{fig:inhale}
\end{figure}

\begin{figure}[htb]
  \centering
  \begin{minipage}{0.4\hsize}
      \includegraphics[keepaspectratio,width=\textwidth]{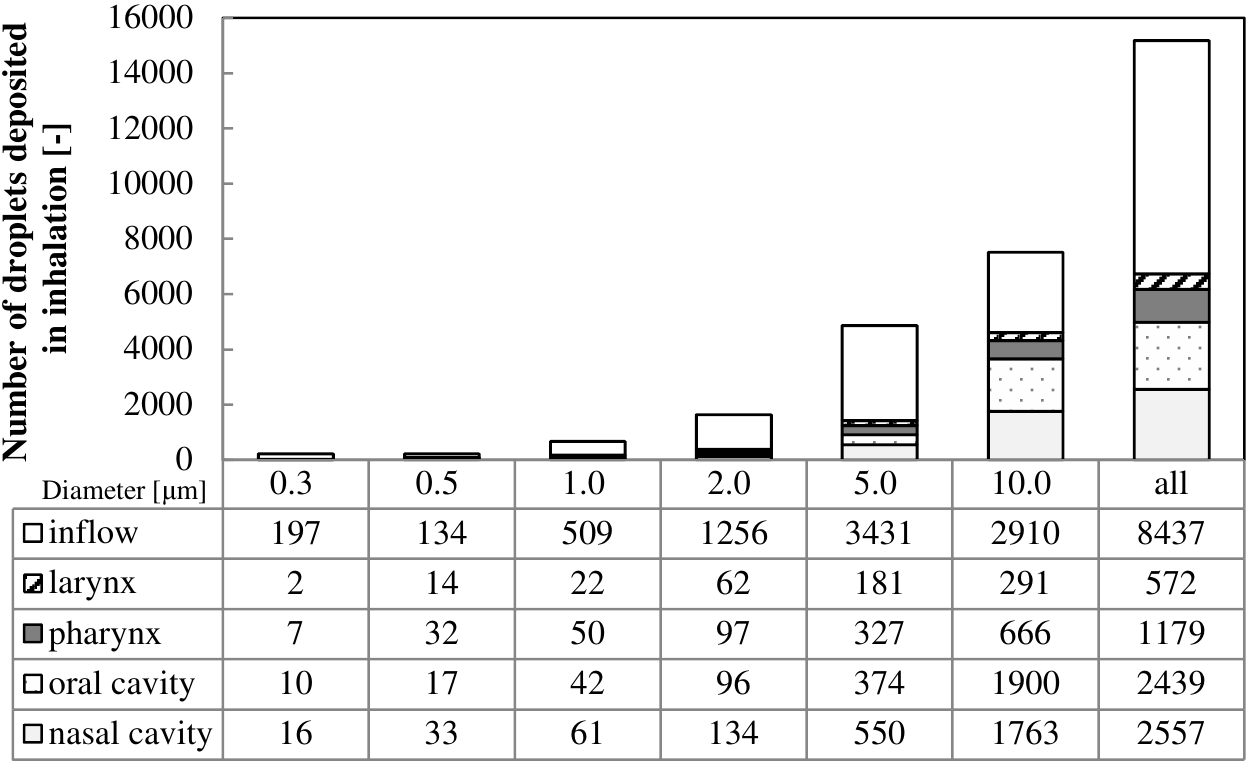}
  \end{minipage}
  \hspace{0cm}
  \begin{minipage}{0.4\hsize}
      \includegraphics[keepaspectratio,width=\textwidth]{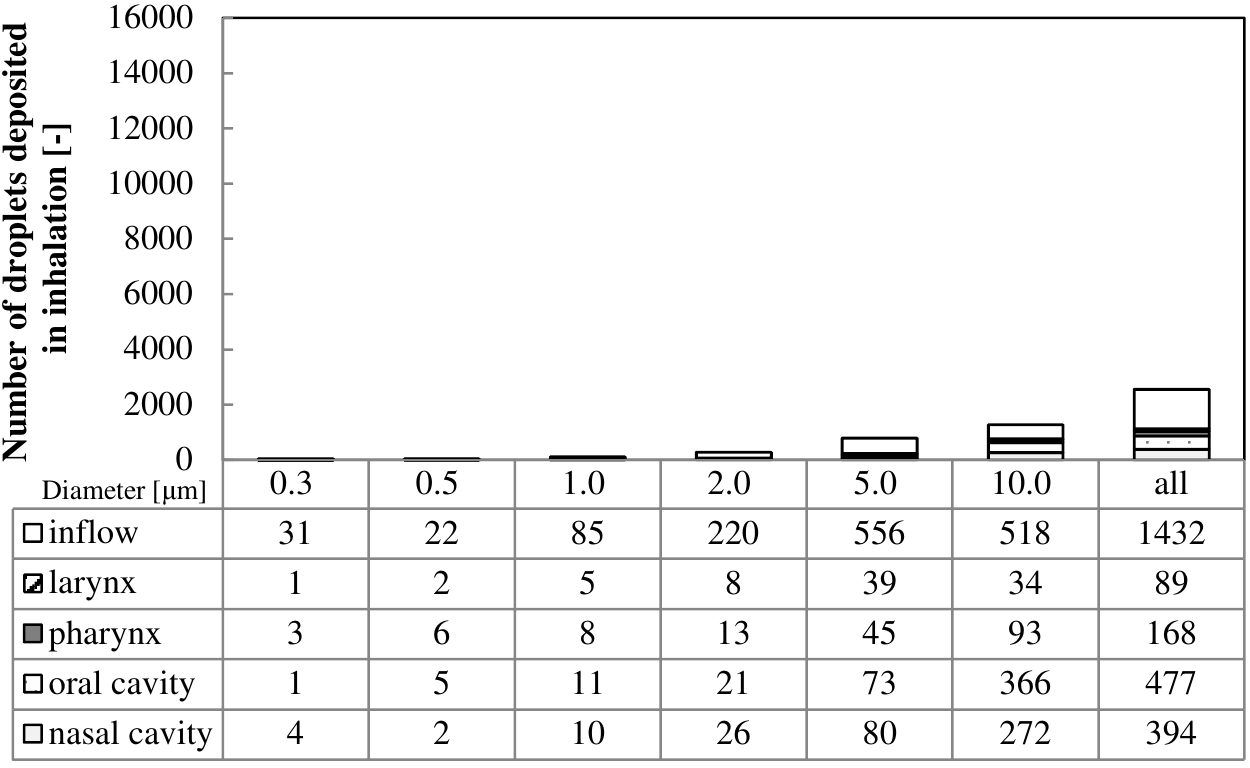}
  \end{minipage}
  \caption{
     Particle deposition rate in each trachea site after 6.0s; left: no mask, right: surgical mask(fit).
  }
  \label{fig:inhale_deposition}
\end{figure}

 The evaluation was performed under the conditions of no mask and surgical mask with a fit state for inhalation protection. Figure \ref{fig:inhale} shows the particle distribution after 6.0 s. Only the particles that have reached near the face are visualized. Figures \ref{fig:inhale_deposition} shows the number of particles deposited at each position of the trachea with and without the mask. Most of the large particles deposited in the nasal and oral cavities were collected by wearing a surgical mask. However, approximately 1,000 particles (reduced by approximately 83\%) reached the depth of the trachea. The number of particles inhaled can be reduced to approximately half by wearing a mask, even when gaps exist. However, it was found that the effect on aerosol particles was limited, and mainly penetrated through the gap between the face and mask.
In addition, the trajectories of the particles reaching the trachea were examined. The particles reached by respiration (for 6.0 s) are within a radius of approximately $0.1 m$ from the center of the mouth opening.

\begin{figure}[htb]
  \centering
  \begin{minipage}{0.7\hsize}
      \includegraphics[keepaspectratio,width=\textwidth]{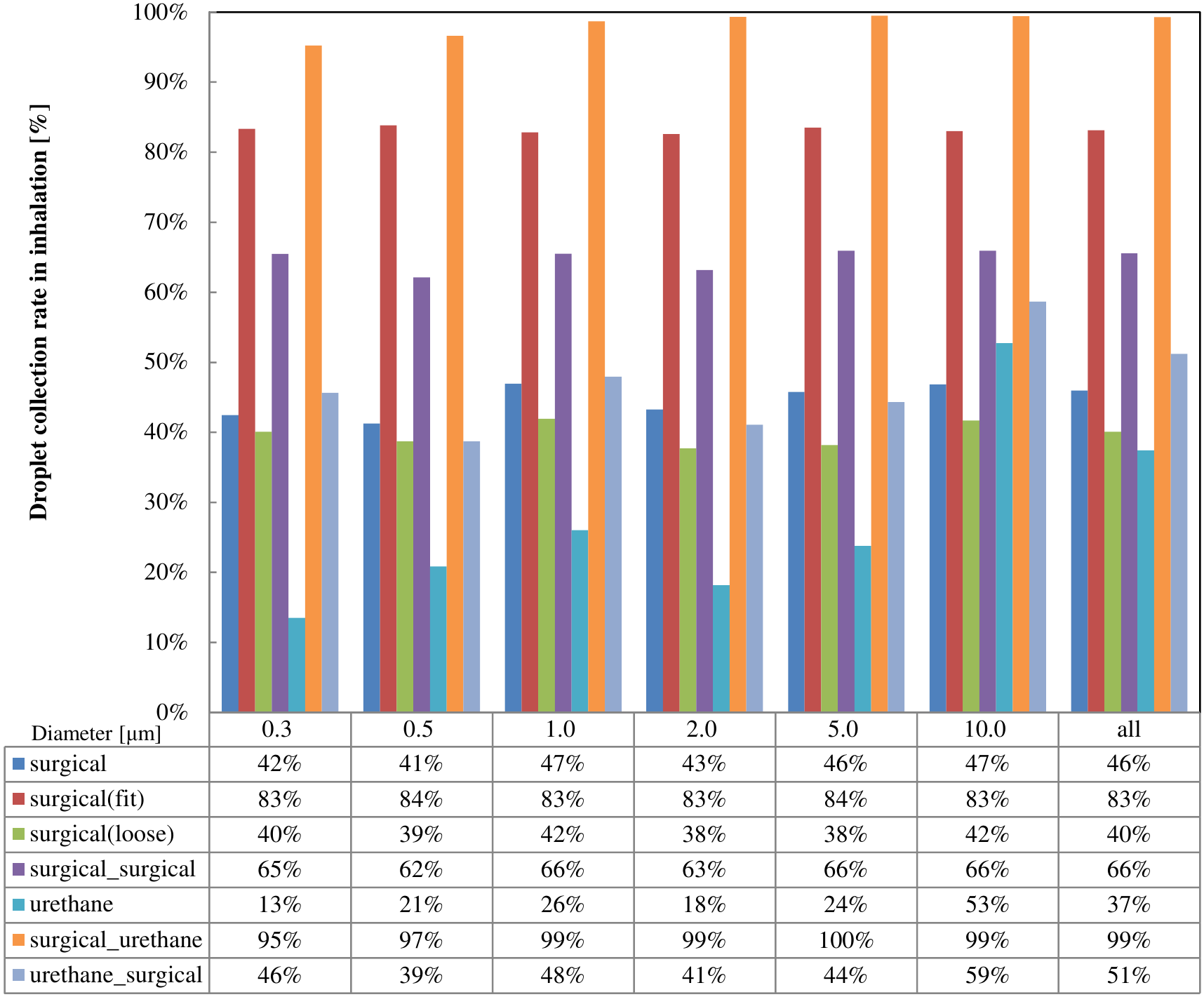}
  \end{minipage}
  \caption{
     Comparison of the collection rates vs. droplet size in double-mask configurations for inhalation protection.
  }
  \label{fig:eff_double_inhale}
\end{figure}

 The performance was evaluated using a double mask. Figure \ref{fig:eff_double_inhale} illustrates the collection rate for each droplet size in double-mask configurations; for instance, \textit{surgical\_urethane} indicates that the surgical and urethane masks are double-mounted.
 In the case of the double surgical mask, no significant improvement in the particle collection performance was observed. Contrary to an intuitive perception, doubling the filter does not improve the performance twice that of a single filter. However, the collection rate is higher when a single mask is in close contact with the face. Hence, the mask performance highly depends on the face-fitting condition. In the case of \textit{surgical\_urethane}, the particle collection rate exceeded 90\%, exhibiting the highest performance among other configurations. Comparable results have been reported in experiments conducted by CDC \cite{CDC2021}, which use a pliable elastomeric source and receiver headforms. This is because the effect of using a second mask to adhere to the face was enhanced rather than the effect of a double filter, as specified in a previous study \cite{CDC2021}. Additionally, there was approximately no improvement in the performance in the case of \textit{urethane\_surgical}, attached in reverse.

\section{Conclusions}

 A 3D flow-droplet multiphase flow analysis on different types of face masks was performed using the topology-free immersed-boundary method. This method enabled us to analyze the effectiveness of face masks on the usage of complex mask shapes and gap areas.

 The following findings were obtained by evaluating the particle collection rate during a single coughing. Both the surgical and fabric cloth masks could collect approximately 60-70\% of the discharged aerosol particles. The surgical mask exhibited a higher performance; however, the cloth mask causes a less leak. In addition, approximately 80\% of the droplets could be collected in both cases considering the volume ratio ($20 \mu m$ or less) index.

 The surgical masks vary depending on the product; however, all products exhibit an overall collection rate of approximately 80\%. The particle images revealed that leakage from the gap beside the nose became dominant depending on how the mask was worn. Therefore, when considering the overall performance of a mask, it is important to consider the performance of the filter and the prevention of particle leakage from the gaps.

 The highest achievable collection rate of aerosol particles in the case of a 1-layered T-shirt mask made of cotton fabric is approximately 30\%. The multi-layered structures improve the collecting performance, although several cases have been overestimated compared to the experimental results. More accurate evaluation might be achieved by considering conditions such as splitting of droplets by the fibers.

 In the case of the face shield, approximately 80\% of the aerosol particles with a size if $5 \mu m$ or less leak. The face shield cannot replace the mask considering the exhalation protection.
 
 In the case of using double masks, two surgical masks or mixed usage of surgical and urethane masks provided an approximately 10 to 15\% in the particle collection performance. This is owing to the narrowing of the gaps between the face and mask by pressing with the outer mask. The collection rate was approximately equivalent to that of a single mask in close contact with the face.

 In the case of inhalation protection, the number of particles can be reduced by approximately half by wearing a surgical mask. In addition, using double surgical masks did not significantly improve the particle collection performance. However, the collection rate was higher when a single mask was in close contact with the face. Again, the mask performance strongly depends on the face-fitting condition. In the case of \textit{surgical\_urethane} mask, the particle collection rate exceeded 90\%, exhibiting the highest performance among all the compared configurations. However, the performance was not improved in the case of \textit{urethane\_surgical} mask, attached in reverse.

 The obtained results as to wearing a mask can be summarized as follows:
\begin{itemize}
\item By wearing the mask correctly, the closing gaps between the face and droplets can be sufficiently suppressed.
\item If a surgical mask is not worn appropriately, more particles may be scattered than those of a cloth mask.
\item The face shield is not a substitute for a face mask to prevent aerosol scattering.
\item Double mask is an appropriate choice; however, it is not as effective as double ``filtering.'' Suppressing the gap between the face and masks is an important factor.
\end{itemize}

 The results of this study have been reported in various media, and we believe it has been an opportunity to deepen people's understanding of face masks. We hope this has contributed to the epidemiological understanding of the reduction of the risk of infection.


%
%

%


\section*{Acknowledgments} 
 This research used computational resources of the supercomputer Fugaku (Evaluation Environment in the trial phase, Project ID:g9330005) provided by the RIKEN Center for Computational Science and the supercomputer Oakbridge-CX (Project ID:hp200154) provided by the University of Tokyo. We would like to thank Editage (www.editage.com) for English language editing.

\section*{Author Declarations} 

\subsection*{Conflict of Interest} 
The authors have no conflicts to disclose.

\subsection*{Data Availability} 
The data that support the findings of this study are available from the corresponding author upon reasonable request.

\subsection*{Authors'' Contributions} 
K.O. carried out all the development of the calculation code and the execution of the simulation. A.I. have worked on building the experimental setup, statistical post-processing of the data, and setting up the mask performance experiment. M.Y. provided all physiological conditions. M.T. managed the research activity.
The authors acknowledge the resources provided by the Ministry of Education, Culture, Sports, Science and Technology.

\bibliography{preprint}

\end{document}